\documentclass[a4paper,11pt]{article}
\pdfoutput=1 
\usepackage{jheppub} 
\usepackage[T1]{fontenc} 
\usepackage{color}

\title{Non-Markovian dynamics of black hole phase transition}

\author[a,b]{Ran Li,}
\author[b,c,*]{Jin Wang \note[*]{Corresponding author}}
\affiliation[a]{School of Physics, Henan Normal University, Xinxiang 453007, China}
\affiliation[b]{Department of Chemistry, Stony Brook University, Stony Brook, NY 11794, USA}
\affiliation[c]{Department of Physics and Astronomy, Stony Brook University, Stony Brook, NY 11794, USA,}

\emailAdd{liran@htu.edu.cn}
\emailAdd{jin.wang.1@stonybrook.edu}

\abstract{We provide a comprehensive study on the non-Markovian dynamics of the black hole phase transitions based on the underlying free energy landscape. In general, there are two typical timescales to characterize non-Markovian dynamics: (i)the timescale of the system represented by the intrinsic oscillating time of the black hole state at the potential well on the free energy landscape; (ii) the correlation time of the effective thermal bath. By assuming that the transition processes between different black hole states are stochastic, the non-Markovian dynamics of the black hole phase transition is governed by the generalized Langevin equation with the time-dependent friction that represents the memory effect from the effective thermal bath when the timescale of the system is comparable or shorter than the timescale of the effective thermal bath. We consider the first passage problem associated with the black hole phase transitions and derive the analytical expressions of the mean first passage time in the weak, intermediate, and large friction regimes. As the concrete examples, we study the effects of three types of time dependent friction kernel (delta function friction, exponentially decayed friction, and oscillatory friction) on the dynamics of Hawking-Page phase transition and the small/large RNAdS black hole phase transition. we found that there is a kinetic turnover point for each type of friction kernel when the friction strength varies. For the exponentially decayed friction kernel, it is shown that the non-Markovian effect slows down the phase transition dynamics in the weak friction regime and speeds up the transition process in the strong friction regime. For the oscillating decayed friction kernel, we found kinetic resonances when the oscillating frequency of the effective thermal bath is equal to the oscillating frequency of the black hole state in the initial potential well on the free energy landscape. }

\begin{document}

\maketitle

\flushbottom

\section{Introduction}

Since the discovery of Hawking radiation \cite{Hawking:1975vcx}, studying the black holes from the viewpoints of thermodynamics and statistical physics has attracted significant attention \cite{Bekenstein:1973ur,Davies:1977bgr,Padmanabhan:2009vy}. It is believed that there is a deep connection between gravity, thermodynamics, and microscopic physics. Black hole phase transition is an important topic in this area \cite{Kubiznak:2014zwa,Kubiznak:2016qmn}, which may provide us insight into the microscopic structures of black holes as well as gravity.

Conventionally, black holes are treated as thermodynamic entities with mass, temperature, and entropy as the general macroscopic objects and the black hole phase transition can be analyzed by calculating the on-shell thermodynamic potential \cite{Gibbons:1976ue} and comparing which phase is thermodynamically favored \cite{Hawking:1982dh}. However, in this circumstance, the dynamics of the black hole phase transition cannot be disclosed, i.e. the dynamical process from one phase to another phase cannot be characterized only by analyzing the on-shell thermodynamic properties.

Recently, the stochastic dynamical models of the black hole phase transition have been proposed to quantify the dynamics of the phase transition process \cite{Li:2020khm,Li:2020nsy,Li:2021vdp}. These models are based on the free energy landscape quantification \cite{FSW,FW,NG,JW}, which can be used to describe the generalized free energy of all the black hole states in the canonical ensemble as the function of the order parameter of the black holes \cite{Li:2020khm,Li:2020nsy}. On the free energy landscape, we are allowed not only to analyze the thermodynamically favored state but also to characterize the dynamical process of the black hole transition from one state to another state. One can refer to references \cite{Wei:2020rcd,Li:2020spm,Wei:2021bwy,Cai:2021sag,Lan:2021crt,Li:2021zep,Yang:2021nwd,Mo:2021jff,Kumara:2021hlt,Li:2021tpu,Liu:2021lmr,Xu:2021usl,Du:2021cxs} for some related works.

However, it should be noted that in these models \cite{Li:2020khm,Li:2020nsy,Li:2021vdp}, only the white noise from the effective thermal bath that represents the microscopic degrees of freedom of the black hole was considered and the corresponding dynamics of the black hole phase transition is Markovian. The Langevin equation based on the free energy landscape was applied to study the first passage problems of the dynamical processes of Hawking-Page phase transition \cite{Hawking:1982dh} and the small/large black hole phase transition \cite{Kubiznak:2012wp}. By calculating the mean first passage time (MFPT) or the Kramers rates of the phase transition process, it is shown that the kinetics of the Reissner-Nordstr\"om Anti-de Sitter (RNAdS) black hole phase transition has a turnover point upon varying the friction coefficient \cite{Li:2021vdp}. In addition, the relative fluctuations of the MFPT were shown to be the monotonic function of the friction coefficient. On the other hand, the friction is assumed to reflect the strength of the interaction of the effective thermal bath from the underlying degrees of freedom of the black hole on the order parameter of the black hole. Therefore, the microstructure of the black hole can be probed by the dynamics of the black hole phase transition. The following study of the dynamics of the Hawking-Page phase transition by taking the Hawking evaporation effect into account shows that there is also a kinetic turnover when Hawking evaporation effect dominates \cite{Li:2021zep}.

In general, there are two typical timescales to characterize the dynamics of the system surrounded by the environment. For our case, they are: (i)the timescale of the system represented by the intrinsic oscillating time of the black hole state at the potential well on the free energy landscape; (ii) the correlation time of the effective thermal bath. In the Markovian model of the black hole phase transition, it was assumed that there is a clear-cut separation between the correlation time related to the effective thermal bath from the underlying degrees of freedom of the black hole and the oscillating time of the black hole state at the initial potential well on the free energy landscape, i.e. the fast correlation time can be ignored when comparing to the oscillating time. In the Markovian models, the friction is instantaneous. If the correlation time of the effective thermal bath is comparable or longer than the oscillating time of the black hole state in the potential well on the free energy landscape, the Markovian treatment of the black hole phase transition is not valid, and the non-Markovian model is required. The non-Markovian effect is then represented by the time-dependent friction, which describes the memory effect from the thermal bath.

In this work, we consider the case where the stochastic noise of the thermal bath is colored, which corresponds to the non-Markovian dynamics in turn. The non-Markovian effect can be represented by the time dependent friction kernel, where the kinetics is then governed by the generalized Langevin equation \cite{Kubo:1985,Nitzan:2006,Zwanzig:1961}. The generalized Langevin equation is an integro-differential equation where the integral term reflects the memory effect or the non-Markovian effect of the effective thermal bath from the underlying degrees of freedom of the black hole. It should be noted that the framework of the non-Markovian model of black hole phase transition has been briefly introduced in \cite{Li:2021}, where the transition rate of the non-Markovian dynamics of black hole phase transition was derived and used to study the kinetics of the Hawking-Page phase transition. It was shown that the non-Markovian effect or the memory effect can promote the dynamical process of Hawking-Page phase transition. However, the conclusion is only valid in the intermediate-strong friction regime.

As studied in the Markovian models of the black hole phase transition, the kinetics of the black hole phase transition based on the non-Markovian model can also be analyzed in the whole friction regime. Previous works showed that the transition rates of the stochastic Brownian particle under the friction represented by the memory kernel are amenable to the analytical derivations in the weak, intermediate, and large friction regimes \cite{Grote:1980II,Grote:1982,Carmeli:1983,Carmeli:1982}. Based on the generalized Langevin equation, the analytical expressions of the MFPTs for the non-Markovian dynamics in the three different friction regimes are derived. It is shown that the analytical expressions of the MFPT in the intermediate friction regime also includes the one applied in the strong friction regime. Note that the transition rate derived in \cite{Li:2021} is only valid in the intermediate-strong friction regime. By employing the analytical expressions of the MFPTs in the intermediate and weak friction regimes, the bridging formula of the MFPT for the black hole phase transition is explicitly given \cite{Carmeli:1984,Schuller:2019ega}.

As the concrete examples, we study the effects of three types of time dependent friction kernels (delta function like, exponentially decayed, and oscillatory friction kernels) on the dynamics of the Hawking-Page phase transition and the small/large RNAdS black hole phase transition. The dependencies of the MFPTs of the black hole phase transitions on the ensemble temperature, the friction coefficient, the decay coefficient, and the oscillating frequency are discussed in details. In this way, the non-Markovian effects on the dynamics of black hole phase transition can be discussed. Similar with the case of the Markovian model, there are kinetic turnover point for each type of friction kernel when the friction strength varies. For the exponentially decayed friction kernel, it is shown that the non-Markovian effect slows down the phase transition dynamics in the weak friction regime and speeds up the transition process in the strong friction regime. For the oscillating decayed friction kernel, there are kinetic resonances when the oscillating frequency of the effective thermal bath is equal to the oscillating frequency of the black hole state in the initial potential well on the free energy landscape.

This paper is arranged as follows. In Sec.\ref{Free_energy}, we briefly review the free energy landscape formalism of the black hole phase transition. In Sec.\ref{Non_Markovian_model}, we describe the non-Markovian dynamics of the black hole phase transition and derive the analytical expressions of the MFPTs in different friction regimes. A bridging formula for the MFPT in the whole friction regime is also discussed. In Sec.\ref{Num_results}, we present the numerical results and the corresponding discussions of the effects of non-Markovian dynamics on the Hawking-Page phase transition and the small/large RNAdS black hole phase transition. The conclusion and discussion are given in Sec.\ref{Con_dis}.

\section{Free energy landscape of the black hole phase transition}
\label{Free_energy}

In this section, we briefly review the free energy landscape formalism of the black hole phase transition, which has been constructed in the previous works \cite{Li:2020khm,Li:2020nsy,Li:2021vdp}. In the following two subsections, two typical phase transitions in black hole physics, i.e., Hawking-Page phase transition and the small/large RNAdS black hole phase transition are considered. 

\subsection{Hawking-Page phase transition}

Without loss of generality, we consider the  Schwarzschild-AdS (SAdS) black hole in four dimensions. The metric is given by \cite{Hawking:1982dh}
\begin{eqnarray}\label{SAdS}
ds^2=-\left(1-\frac{2M}{r}+\frac{r^2}{L^2}\right)dt^2+
\left(1-\frac{2M}{r}+\frac{r^2}{L^2}\right)^{-1}dr^2+r^2d\Omega_2^2\;,
\end{eqnarray}
where $M$ and $L$ are the black hole mass and the AdS curvature radius. When $M=0$, the metric (\ref{SAdS}) gives the metric of pure AdS space in four dimensions. The mass, the Hawking temperature, and the Bekenstein-Hawking entropy of the SAdS black hole are respectively given by
\begin{eqnarray}
M&=&\frac{r_+}{2}\left(1+\frac{r_+^2}{L^2}\right) \;,
\\
T_H&=&\frac{1}{4\pi r_+}\left(1+\frac{3r_+^2}{L^2}\right)\;,\\
S&=&\pi r_+^2\;,
\end{eqnarray}
where $r_+$ is the black hole radius. The black hole radius is determined by the equation $1-\frac{2M}{r}+\frac{r^2}{L^2}=0$ and is given by the largest root of this equation.

Besides the thermal AdS space and the small/large SAdS black holes, which are the solutions to the Einstein equations, we also assume that there are black holes with arbitrary horizon radii generated by the thermal fluctuations. These black holes are the fluctuating black holes. Following the discussion in \cite{Li:2020khm}, we consider the canonical ensemble at the specific temperature $T$, which is composed of a series of black hole spacetimes with an arbitrary horizon radius. We call the black hole spacetime in the ensemble as the black hole state. The black hole state in the ensemble is then characterized by the black hole radius, which is considered as the order parameter. According to the ensemble theory, every black hole state in the ensemble has the specific probability, which is determined by the generalized free energy function. Our aim is to construct the relationship between the generalized free energy and the order parameter. 

Our starting point is the on-shell Gibbs free energy of the SAdS black hole. According to the Euclidean gravitational path integral \cite{Gibbons:1976ue}, the on-shell Gibbs free energy for the black hole solution to the Einstein equation can be calculated directly from the Euclidean Einstein-Hilbert action, or from  the thermodynamic relationship $G=M-T_HS$. In order to quantify the free energy landscape, we need to specify every spacetime state in the ensemble a Gibbs free energy. We generalize the on-shell Gibbs free energy to the off-shell Gibbs free energy by replacing the Hawking temperature $T_H$ with the ensemble temperature $T$ in the thermodynamic relationship, which is explicitly given as follows \cite{York:1986it,Spallucci:2013osa,Andre:2020czm,Andre:2021ctu}
\begin{eqnarray}
G=M-TS=\frac{r_+}{2}\left(1+\frac{r_+^2}{L^2}\right)-\pi T r_+^2\;.
\end{eqnarray}
In the above equation, the black hole radius $r_+$ is treated as the order parameter, which is emergent from the underlying microscopic degrees of freedom of the black hole. The order parameter $r_+$ in the above equation changes continuously. The generalized free energy is then the continuous function of the order parameter and ensemble temperature.

\begin{figure}
  \centering
  \includegraphics[width=6cm]{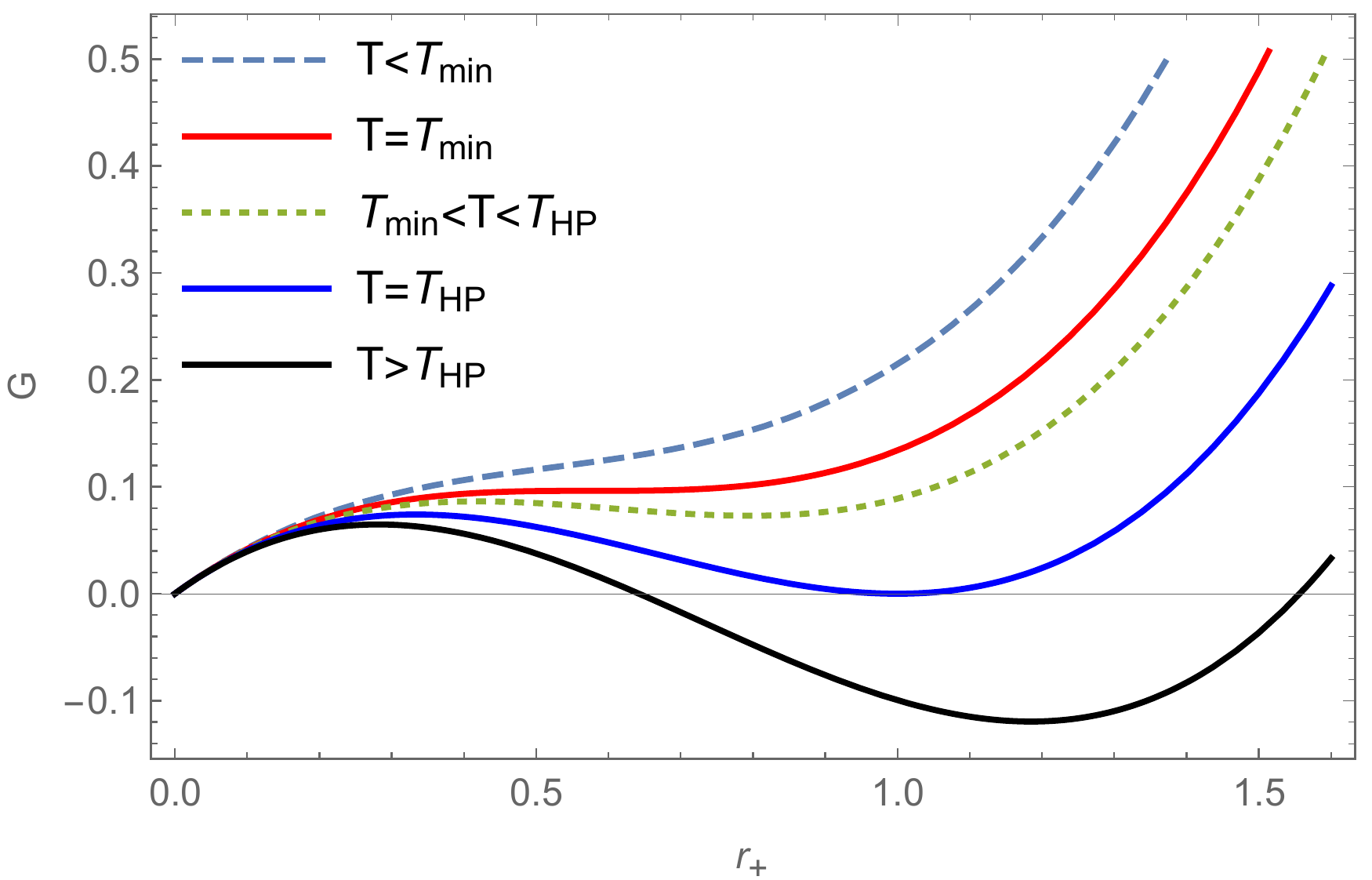}
  \caption{Free energy landscapes for Hawking-Page phase transition. The generalized Gibbs free energy is plotted as the function of black hole radius $r_+$ at different temperature. In discussing Hawking-Page phase transition, we set the AdS curvature radius $L$ as unity. }
  \label{GibbsPlot}
\end{figure}

In Figure \ref{GibbsPlot}, the free energy landscapes for the Hawking-Page phase transition are plotted for different ensemble temperatures. The shape of the free energy landscape for the Hawking-Page phase transition is only modulated by the ensemble temperature $T$. In these plots, the origin $r_+=0$ represents the pure radiation phase or thermal AdS space. When $T<T_{min}=\frac{\sqrt{3}}{2\pi L}$, the free energy is the monotonic increasing function of the order parameter and the origin $r_+=0$ is the global minimum on the free energy landscape. The system has only one stable phase, i.e. the thermal AdS space. At $T=T_{min}$, Gibbs free energy landscape exhibits an inflection point at $r_+=L/\sqrt{3}$. From this temperature on, the two black hole phases emerge (large and small black holes) with radii given by
\begin{eqnarray}\label{rls}
r_{l,s}=\frac{T}{2\pi T_{min}^2}\left(1\pm\sqrt{1-\frac{T_{min}^2}{T^2}}\right)\;.
\end{eqnarray}
The small/large SAdS black hole phase is represented by the maximum/minimum point on the free energy landscape. The small black hole phase is always unstable while the large black hole phase can be stable or unstable depending on the temperature.

The stability of the large SAdS black hole can be analyzed in terms of the fact that the global or the local minimum on the free energy landscape corresponds to the global or the local stable phase. From the plots of the free energy landscapes, we can easily observe that below the Hawking-Page critical temperature $T=T_{HP}=\frac{1}{\pi L}$, the thermal AdS phase is thermodynamically stable, and above the critical temperature, the large black hole phase is thermodynamically stable. At the critical temperature, both the thermal AdS space phase and the large black hole phase are stable with equal free energy basin depth. Since there is a discontinuous change in the order parameter from the thermal AdS phase $r_+=0$ to the large black hole phase $r_+=r_l$ at the Hawking-Page critical temperature, the associated derivative of the free energy function will be divergent, which is a signature of the first order phase transition. It should be noted that besides the thermal AdS space and the small and large SAdS black holes represented by the maximum or the minimum points on the free energy landscape, there are fluctuating black holes on the free energy landscape. The fluctuating black holes are the intermediate states of the phase transition.

\subsection{RNAdS black hole phase transition}

In the present subsection, we consider another example of the black hole phase transition based on the free energy landscape, i.e. the small/large RNAdS black hole phase transition \cite{Kubiznak:2012wp}. The metric of RNAdS black hole in four dimensions is given by 
\begin{eqnarray}
ds^2=-\left(1-\frac{2M}{r}+\frac{Q^2}{L^2}+\frac{r^2}{L^2}\right)dt^2+
\left(1-\frac{2M}{r}+\frac{Q^2}{L^2}+\frac{r^2}{L^2}\right)^{-1}dr^2+r^2d\Omega_2^2\;,
\end{eqnarray}
where $M$ is the mass, $Q$ is the electric charge, and $L$ is the AdS curvature radius.

The mass, the Hawking temperature, and the Bekenstein-Hawking entropy of RNAdS black hole are given by
\begin{eqnarray}
M&=&\frac{r_+}{2}\left(1+\frac{r_+^2}{L^2}+\frac{Q^2}{L^2}\right)\;,\\
T_H&=&\frac{1}{4\pi r_+}\left(1+\frac{3r_+^2}{L^2}-\frac{Q^2}{L^2}\right)\;,\\
S&=&\pi r_+^2\;,
\end{eqnarray}
where $r_+$ is the black hole radius.

Analogous to the case of Hawking-Page phase transition, we consider the canonical ensemble at the specific temperature $T$, which is constructed by including the three branches of the RNAdS black holes (the small, the large, and the intermediate black holes) as well as the fluctuating black holes \cite{Li:2020nsy,Li:2021vdp}. The black holes in the ensemble are also distinguished or described by the continuous order parameter, the horizon radius of the black hole. The three branches of the RNAdS black hole solutions are the solutions to the Einstein field equations, while the fluctuating black holes, which are generated by the thermal noise, are not necessarily the solutions to the Einstein field equations.

The on-shell Gibbs free energy of the three branches of the RNAdS black hole calculated from the Einstein-Hilbert action can be properly rewritten as the thermodynamic relationship of $G=M-T_H S$. We define the generalized off-shell Gibbs free energy of the fluctuating black hole as \cite{York:1986it,Spallucci:2013osa,Andre:2020czm,Andre:2021ctu}
\begin{eqnarray}\label{GibbsEq}
G=M-TS=\frac{r_+}{2}\left(1+\frac{8}{3}\pi P r_+^2+\frac{Q^2}{r_+^2} \right)-\pi T r_+^2\;,
\end{eqnarray}
where $T$ is the ensemble temperature, $r_+$ is the radius of the fluctuating black hole or the order parameter, $P=-\frac{\Lambda}{8\pi}=\frac{3}{8\pi}\frac{1}{L^2}$ is an effective thermodynamic pressure with $\Lambda$ being the cosmological constant (negative value corresponding to AdS universe) and $L$ being the AdS curvature radius, and $Q$ is the charge of the black hole. Note that contrary to the case of Hawking-Page phase transition, we do not set the AdS radius $L$ to unity but treat it as the thermodynamic pressure $P$, which can be adjusted in the following discussion.

\begin{figure}
  \centering
  \includegraphics[width=6cm]{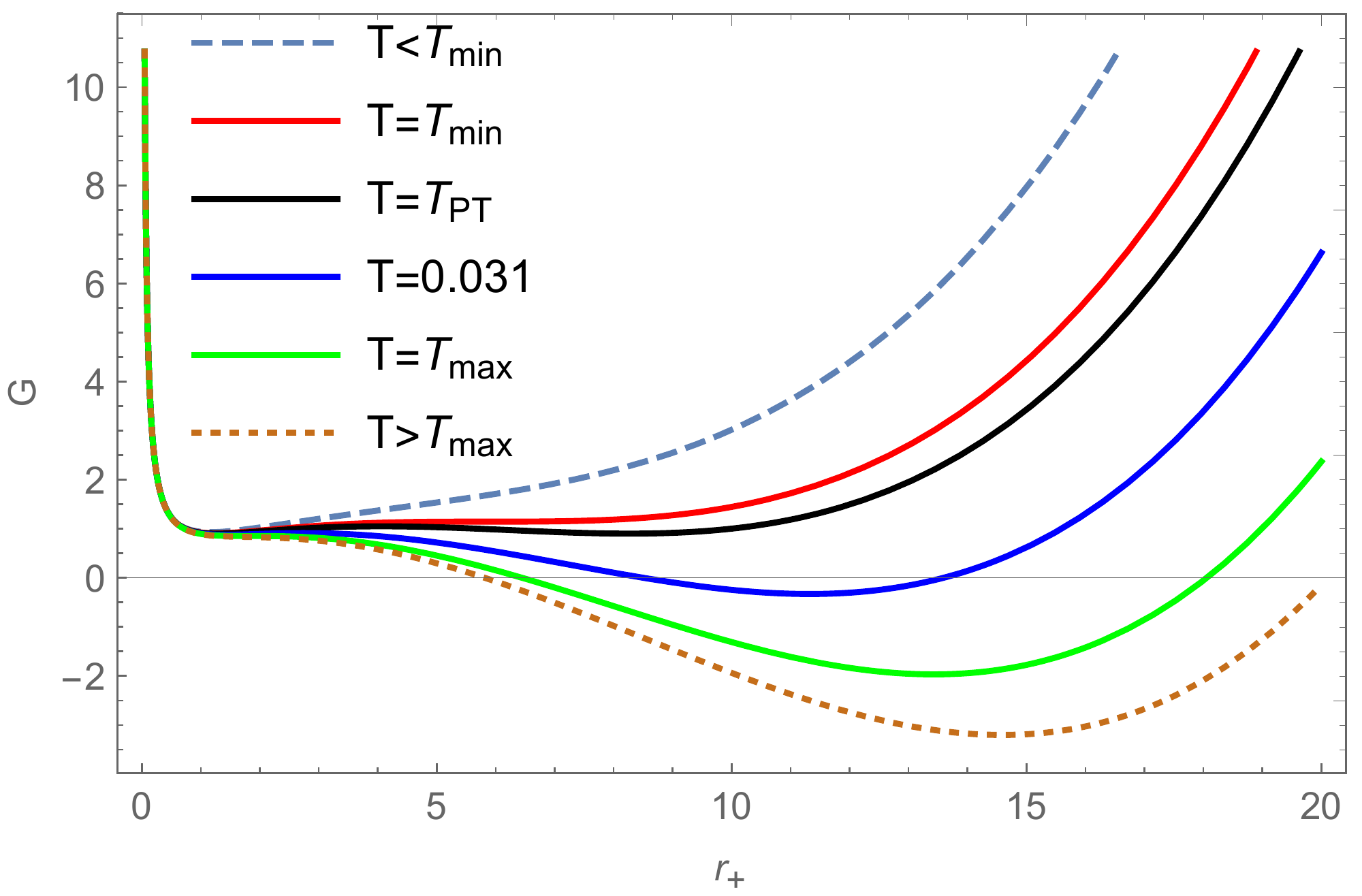}\\
  \caption{Free energy landscapes for the RNAdS black holes. The generalized Gibbs free energy given by Eq.(\ref{GibbsEq}) is plotted as the function of black hole radius $r_+$ for $P<P_c=\frac{1}{96\pi Q^2}$. In the plot, $P=0.32 P_c$, $Q=1$, $T_{min}=0.0256$, $T_{max}=0.0344$ and $T_{PT}=0.02703$ is the phase transition temperature. $P_c$ is the critical pressure.}
  \label{GPlot_diffT}
\end{figure}

We quantify the free energy landscape of the RNAdS black hole by plotting the generalized Gibbs free energy as a function of black hole radius $r_+$ in Fig.\ref{GPlot_diffT}. It is shown that the shape of the free energy function varies along with the ensemble temperature. The detailed analysis of the free energy landscape topography can give rise to the phase diagram of the RNAdS black hole, which gives us all the information about the phase transition.

It can be observed that there exists two temperatures $T_{min}$ and $T_{max}$. When $T<T_{min}$ or $T>T_{max}$, the free energy landscape only has a single well. In this case, there is the small black hole phase or the large black hole phase respectively. When $T_{min}<T<T_{max}$, the free energy landscape exhibits the shape of the double well. The small and the large black holes are locally or globally stable states, while the fluctuating black holes and the intermediate black hole are unstable states. At the critical temperature $T=T_{PT}$, the small black hole phase and the large black hole phase have the equal free energy basin depth. This is the signature of the small/large black hole phase transition. On the free energy landscape, the unstable black holes are treated as the transient states, which are the bridges in the phase transition process. The probability of generating a fluctuating black hole is then determined by the generalized Gibbs free energy via the Boltzmann law $p(r_+)\sim e^{-G(r_+)/k_BT}$. In this sense, the generalized Gibbs free energy can be taken as the effective potential when studying the dynamics of the black hole phase transition. This is the free energy landscape description of the RNAdS black holes. This description is universal in studying the black hole phase transition \cite{Wei:2020rcd,Li:2020spm,Wei:2021bwy,Cai:2021sag,Lan:2021crt,Li:2021zep,Yang:2021nwd,Mo:2021jff,Kumara:2021hlt,Li:2021tpu,Liu:2021lmr,Xu:2021usl,Du:2021cxs}.

\section{Non-Markovian dynamics of the black hole phase transition} \label{Non_Markovian_model}

Recently, the stochastic dynamical model of the black hole phase transition have been proposed to quantify the kinetics of the phase transition process \cite{Li:2020khm,Li:2020nsy,Li:2021vdp}. However, it should be noted that in this model, the white noise from the thermal bath corresponding to the Markovian dynamics is considered, where the Langevin equation based on the free energy landscape has been applied to investigate the black hole phase transition. As discussed in the introduction, if the correlation time related to the effective thermal bath is comparable or even longer than the oscillating time of the black hole state at the initial potential well on the free energy landscape, the Markovian treatment of the black hole phase transition is not valid, and the non-Markovian model is required to study the memory effect. In this section, we introduce the basic concepts and models of the non-Markovian dynamics with specific applications to black hole phase transition.

\subsection{Non-Markovian model}

In the previous Markovian models of the black hole phase transition based on the free energy landscape, it was assumed that the dynamics is govern by the Langevin equation \cite{Li:2021vdp}. The Langevin equation that describes the evolution along the order parameter is considered to be the effective equation emergent from the microscopic degrees of freedom of the black hole. The basic idea comes from the Mori-Zwanzig formalism \cite{Mori:1965,Zwanzig:1960jcp}.

Macroscopic system with a large number of microscopic degrees of freedom such as black hole can be well described by a small number of relevant variables. By applying the Mori-Zwanzig projection, one can find the macroscopic equations that only depend on the relevant variables from the microscopic equations of motion of a system. One can refer to \cite{Vrugt:2021sfu} for the recent application of the Mori-Zwanzig formalism in the averaging problem in cosmology. The irrelevant microscopic degrees of freedom of black hole can be taken to be the effective thermal bath. In the generalized Langevin equation, it is assumed that the dynamics of the black hole phase transition is determined by the three types of forces. The first is the thermodynamic driving force determined by the generalized free energy function. The second force is the effective friction along the order parameter, which is interpreted as the interaction or the dissipation of the microscopic degrees of freedom from the effective heat bath acting on the black hole collective order parameter. The third force represents the stochastic force that comes from the microscopic degrees of freedom of the effective heat bath on the macroscopic order parameter.

In studying the non-Markovian model of black hole phase transition, the correlation time of the thermal bath is assumed to be comparable or even longer than the oscillating time of the black hole state at the potential well on the free energy landscape. The non-Markovian effect can be described by the time dependent friction kernel. Therefore, the non-Markovian dynamics can be described by the generalized Langevin equation complemented by the free energy potential as \cite{Mori:1965,Kubo:1966}
\begin{eqnarray}\label{Non_Mar_Eq}
&&\frac{dr}{dt}=v\;,\nonumber\\
&&\frac{dv}{dt}=-\frac{dG(r)}{dr}-\int_{0}^{t} \zeta(\tau) v(t-\tau) d\tau
+\eta(t)\;,
\end{eqnarray}
where the centered color noise $\eta(t)$ with zero mean and the dissipation kernel $\zeta(t)$ are related by the second fluctuation-dissipation theorem 
\begin{eqnarray}
\zeta(|t|)=\frac{1}{k_B T} \langle \eta(0)\eta(t)\rangle\;.
\end{eqnarray}
In Eq.(\ref{Non_Mar_Eq}), we have abbreviated the order parameter $r_+$ to $r$ for simplicity. As noted, the Langevin equation with the Markovian dynamics in \cite{Li:2021vdp} can be reduced from Eq.(\ref{Non_Mar_Eq}) by taking the friction kernel instantaneous, i.e. $\zeta(t)=\zeta\delta(t)$. The non-Markovian effects is reflected by the time dependent friction kernel $\zeta(t)$. The generalized Langevin equation can be viewed as the effective equation that describes the evolution of the black hole order parameter at the macroscopic emergent level. It is an integro-differential equation.

Now, we will study the dynamics of black hole phase transition by treating the process as the first passage problem. First passage time is a very important quantity in characterizing the speed of the dynamics, which can be defined as the time required for one stable black hole state crossing the intermediate transient state to reach another stable state for the first time. On the free energy landscape, this problem can be regarded as the black hole state in the initial potential well crossing the barrier in the middle to reach the potential well on the other side. However, because of the stochastic nature of the transition process described by the Langevin equation, the first passage time is a stochastic quantity. The mean first passage time (MFPT) is then defined as the average timescale for a stochastic event to first occur. The MFPT is the characteristic time that describes the dynamics of the phase transition process.

\subsection{MFPT: large friction}

In this section, we will derive the analytical expressions of the MFPT for the large friction case. Note that in deriving the analytical expressions of the MFPTs, we have assumed that the free energy landscape has the shape of double well as in the case of the RNAdS black holes. We mainly concern the transition rate or the MFPT of the dynamical transition process from the state in the left well to the state in the right well. When considering the Hawking-Page phase transition, the final results can be easily obtained by replacing the corresponding quantities in the RNAdS case with the corresponding ones in the Hawking-Page phase transition case.

In the large friction limit, the corresponding non-Markovian Fokker-Planck equation is in the form \cite{Sokolov:2002,Assis:2006}
\begin{eqnarray}\label{non_markovian_FPeq}
\frac{\partial}{\partial t}\rho_(r, t) =\int D(t-\tau) \mathcal{L} \rho(r, t) d\tau
\;,
\end{eqnarray}
where $D(t)$ is the diffusion kernel and $\mathcal{L}$ is the linear operator acting on the order parameter $r$. The operator $\mathcal{L}$ has the same form as the Markovian model 
\begin{eqnarray}
\mathcal{L}= \frac{\partial}{\partial r} e^{-\beta G(r)} \frac{\partial}{\partial r} e^{\beta G(r)} \;,
\end{eqnarray}
where $\beta=1/k_B T$ is the inverse temperature. It should be noted that the diffusion kernel and the friction kernel have the relation as
\begin{eqnarray}\label{D_zeta_relation}
\int_0^t D(\tau)\zeta(t-\tau) d\tau=k_B T \delta(t)\;.
\end{eqnarray}

The non-Markovian Fokker-Planck equation describes the time evolution of the probability distribution $\rho(r,t)$ in the ensemble. It is an integro-differential equation, which is obviously harder to deal with than the Fokker-Planck equation for the Markovian model. The non-Markovian Fokker-Planck equation (\ref{non_markovian_FPeq}) with a $\delta$-functional diffusion kernel $D(t)=D\delta(t)$ where $D=\frac{k_B T}{\zeta}$ corresponds to a usual Fokker-Planck equation describing the Markovian processes.

To calculate the MFPT, our starting point is the non-Markovian Fokker-Planck equation (\ref{non_markovian_FPeq}). However, as mentioned, it is difficult to deal with by using both the analytical and numerical methods. Firstly, we employ the Laplace transformation of the non-Markovian Fokker-Planck equation, which can be written in the frequency space as \cite{Bryngelson:1989}
\begin{eqnarray}
s\rho(r,s)-n_i(r)= \hat{D}(s) \frac{\partial}{\partial r} e^{-\beta G(r)} \frac{\partial}{\partial r} e^{\beta G(r)} \rho(r,s)\;.
\end{eqnarray}
This equation is similar to the Fokker-Planck equation in the frequency space that have been considered in \cite{Li:2020khm} except the diffusion coefficient $D$ is replaced by the frequency dependent diffusion kernel $\hat{D}(s)$. Therefore, the frequency dependent diffusion kernel $\hat{D}(s)$ includes all the non-Markovian effects on the dynamics. From the relation between the diffusion kernel and the friction kernel, i.e., Eq.(\ref{D_zeta_relation}), the Laplace transformations give the relation in the frequency space as 
\begin{eqnarray}
\hat{D}(s) \hat{\zeta}(s) =k_B T\;,
\end{eqnarray}
which is analogous to the fluctuation-dissipation relation for the Markovian dynamics, while this relation is valid in the frequency space.

Defining $\Sigma(t)=\int_{0}^{r_l}\rho(r,t)dr$ as the probability that the state has not made a first passage by time $t$. Note that the first passage time is a random variable. The distribution of the first passage time is then given by $F_{p}(t)=-\frac{d\Sigma(t)}{dt}$. The MFPT defined as the average of the first passage time is then given by 
\begin{eqnarray}\label{MFPT_def}
\hat{t}=\int_{0}^{+\infty} t F_p(t) dt= \int_{0}^{r_l} \rho(r,s=0) dr\;.
\end{eqnarray}

Following the same procedure as done in \cite{Li:2020khm}, one can derive the relation from the Fokker-Planck equation in frequency space \begin{eqnarray}
\rho(r,s=0)=\frac{1}{\hat{D}(0)}\int_{r}^{r_l} dr'\int_{0}^{r'} dr'' n_i(r'')   e^{\beta\left(G(r')-G(r)\right)}\;.
\end{eqnarray}
In deriving this equation, we have used the reflecting boundary condition at $r=0$ and the absorbing boundary condition at $r=r_l$. Finally, by substituting this equation into Eq.(\ref{MFPT_def}) and exchanging the order of integration, one can obtain the analytical integration expression for the MFPT of the transition process as \cite{Bryngelson:1989}
\begin{eqnarray}\label{MFPT_int}
\bar{t}=\frac{1}{\hat{D}(0)} \int_{r_i}^{r_l} dr 
\int_{0}^{r} dr' e^{\beta\left(G(r)-G(r')\right)} \;,
\end{eqnarray}
where $\hat{D}(0)$ is the zero frequency value of the frequency dependent diffusion kernel. In deriving this equation, the initial distribution $n_i(r)$ is selected to be $\delta(r-r_i)$, where $r_i$ is taken to be $r_i=r_s$ representing the small RNAdS black hole for the small/large black hole phase transition. Then, Eq.(\ref{MFPT_int}) can be interpreted as the MFPT of the transition process from the small to the large RNAdS black hole phase. The MFPT for the inverse process can also be derived analogously \cite{Li:2020khm}. 

For the Hawking-Page phase transition, we only consider the transition process from the large SAdS black hole state in the right well to the thermal AdS state at $r=0$. Therefore, the reflecting boundary condition is imposed at $r=+\infty$ and the absorbing boundary condition is imposed at $r=0$. The expression of MFPT for this process is similar to the expression of MFPT for RNAdS case from the large black hole phase to the small black hole phase \cite{Li:2020khm}.

There is an approximate formula for the MFPT in the large friction regime. Note that the Gibbs free energy near the minimum and the maximum can be approximated by the quadratic expansion as 
\begin{eqnarray}\label{G_app}
G(r)&=&G(r_s)+\frac{1}{2}\omega_{s}(r-r_s)^2+\cdots\nonumber\\\;,
G(r)&=&G(r_m)+\frac{1}{2}\omega_{m}(r-r_m)^2+\cdots\;.
\end{eqnarray}
By performing the Gaussian integral, the approximate expression of the MFPT in the high barrier limit $(\beta W\gg 1)$ can be obtained as 
\begin{eqnarray}\label{MFPT_large_exp}
\bar{t}=\frac{2\pi\hat{\zeta}(0)}{\omega_s\omega_m} e^{\beta W}\;.
\end{eqnarray}
This formula of the MFPT is only valid for the case of large friction and high barrier. Note that this expression is valid for the RNAdS black hole phase transition. For the Hawking-Page phase transition, $\omega_{s}$ and $\omega_{m}$ should be replaced by $\omega_{l}$ and $\omega_{s}$ for the free energy landscape of SAdS black holes, and $W$ is the barrier height between the unsatble small SAdS black hole state and the large SAdS black hole state on the free energy landscape.

\subsection{MFPT: intermediate friction}

In the intermediate friction regime, the corresponding generalized Fokker-Planck equations for the Brownian oscillators were derived in \cite{Adelman:1976}. Note that the potential function used to derive the generalized Fokker-Planck equations is quadratic. The starting point is the probability distribution $\rho(r,v,t|r_0,v_0)$ with the initial parameters $r_0$ and $v_0$. Since the random force $\eta(t)$ is assumed to be Gaussian, the probability distribution $\rho(r,v,t|r_0,v_0)$ for the Brownian oscillators is shown to be also Gaussian, which is given by \cite{Adelman:1976} 
\begin{eqnarray}
\rho(r,v,t|r_0,v_0)=\frac{1}{\pi}\left|\det{Q^{-1}} \right|^{1/2}
\exp(-y^T\cdot Q^{-1}\cdot y)\;, 
\end{eqnarray}
where 
\begin{eqnarray}
y^T&=&(r(t)-\langle r(t) \rangle,v(t)-\langle v(t) \rangle)\;,
\nonumber\\
Q_{ij}&=&2\langle y_i(t) y_j(t) \rangle\;. 
\end{eqnarray}
The averages in the above equations should be taken over an evolving swarm of trajectories that accounts for the initial conditions $r(0)=r_0=-a$ and $v(0)=v_0$. In terms of Laplace transformation of the generalized Langevin equation (\ref{Non_Mar_Eq}), the time evolution equations for $Q_{ij}$ can be calculated. Adelman showed that the probability distribution function $\rho(r,v,t)$ satisfies the generalized Fokker-Planck equation \cite{Adelman:1976}.

For our case, the free energy function near the top of the barrier $r=r_m$ can also be approximated as quadratic function as Eq.(\ref{G_app}). Therefore, the generalized Fokker-Planck equation around the top of the potential barrier $r=r_m$ can be given by \cite{Adelman:1976} 
\begin{eqnarray}\label{GFPE}
\frac{\partial}{\partial t} \rho(r,v,t) &=&\left[
-v\frac{\partial }{\partial r} 
-\hat{\omega}_m^2 r \frac{\partial }{\partial v} +\gamma(t)\left( \frac{\partial}{\partial v} v + k_B T \frac{\partial^2}{\partial v^2}\right)
\right.\nonumber\\
&&\left.
+k_B T \left(\frac{\hat{\omega}_m^2}{\omega_m^2}-1\right)\frac{\partial^2}{\partial r\partial v}
\right]\rho(r,v,t)\;,
\end{eqnarray}
with 
\begin{eqnarray}
&&\gamma(t)=-\frac{d}{dt}\ln\Phi(t)\;,\nonumber\\
&&\hat{\omega}_m^2(t)=-\theta(t)/\Phi(t)\;,\nonumber\\
&&\Phi(t)=\dot{\chi}(t)\left[1+\omega_m^2 \int_0^t d\tau \chi(\tau)\right] -\omega_m^2 \chi^2(t)\;,\nonumber\\
&&\theta(t)=\omega_m^2\left[\chi(t)\ddot{\chi}(t)
-\dot{\chi}^2(t)\right]\;,
\end{eqnarray}
and the function $\chi(t)$ is defined as the inverse Laplace transformation 
\begin{eqnarray}
\chi(t)=\mathcal{L}^{-1}\left[\frac{1}{s^2-\omega_m^2+\hat{\zeta}(s) s}\right]\;.
\end{eqnarray}
For a detailed derivation, one can refer to \cite{Adelman:1976,Hanggi:1982}. Comparing the Fokker-Planck equation for Markovian noise \cite{Li:2021vdp} with the generalized Fokker-Planck equation (\ref{GFPE}), several similarities are remarkable. In Eq.(\ref{GFPE}), the last term is an additional diffusive term that is absent in the Markovian case. The other terms can also be found in the Markovian case. However, $\gamma(t)$ is time dependent while the corresponding coefficient in the Markovian case is a constant.

Next, one can find the approximate steady-state solution to the generalized Fokker-Planck equation (\ref{GFPE}). By inserting the ansatz \cite{Carmeli:1984}
\begin{eqnarray}
\rho(r,v,t)=\frac{1}{Z}\Theta(r,v,t) e^{-\beta\left(v^2/2+G(r)\right)}\;,
\end{eqnarray}
into the generalized Fokker-Planck equation (\ref{GFPE}), and assuming the function $\Theta(r,v,t)$ of the form 
\begin{eqnarray}
\Theta(r,v,t)=\Theta(u,t)\;,\;\;u=v+\Gamma r\;,
\end{eqnarray}
one can derive the resultant equation for $\Theta$ as given by 
\begin{eqnarray}
\frac{\partial \Theta}{\partial t}=-k_B T (\bar{\lambda}+\Gamma)\frac{\partial^2 \Theta}{\partial  u^2}+\bar{\lambda}\left[v-\frac{\omega_m^2}{\bar{\lambda}}r\right] \frac{\partial \Theta}{\partial u}\;,
\end{eqnarray}
where 
\begin{eqnarray}
\bar{\lambda}=-\left[\gamma(t)+\Gamma\frac{\bar{\omega}_m^2(t)}{\omega_m^2}\right]\;.
\end{eqnarray}

By solving the resultant equation with the appropriate boundary conditions, one can derive the steady-state solution of $\Theta$ as \cite{Schuller:2019ega}
\begin{eqnarray}
\Theta(u)=\sqrt{\frac{\lambda}{2\pi A}}\int_{-\infty}^u\exp\left[-\frac{\lambda s^2}{2A}\right]ds\;,
\end{eqnarray}
where 
\begin{eqnarray}
&&A=k_BT \left(\gamma-b c\right)\;,\nonumber\\
&&b=\frac{\omega_m^2}{\bar{\omega}_m^2}\left(\frac{\gamma}{2}+\sqrt{\frac{\gamma^2}{4}+\bar{\omega}_m^2}\right)\;,\nonumber\\
&&c=\frac{\bar{\omega}_m^2}{\omega_m^2}-1\;,\nonumber\\
&&\gamma=\lim_{t\rightarrow\infty}\gamma(t)\;,\nonumber\\
&&\bar{\omega}_m^2=\lim_{t\rightarrow\infty}\bar{\omega}_m^2(t)\;.\nonumber
\end{eqnarray}
and the reactive frequency $\lambda$ is determined by the equation \cite{Peters:2017}
\begin{eqnarray}\label{lambda_eq}
\lambda=\frac{\omega_m^2}{\lambda+\hat{\zeta}(\lambda)}\;.
\end{eqnarray}
Note that the frequency component of the time dependent friction $\hat{\zeta}(\lambda)$ is the inverse Laplace transformation of the friction kernel $\zeta(t)$
\begin{eqnarray}
\hat{\zeta}(\lambda)=\int_{0}^{\infty} e^{-\lambda t} \zeta(t) dt\;.
\end{eqnarray}
The entire information about the friction kernel $\zeta(t)$ is therefore completely contained in $\lambda$. In practice, $\lambda$ is the largest root of the equation (\ref{lambda_eq}).

Finally, one can compute the transition rate from the small black hole state to the large black hole state, which is explicitly given by \cite{Grote:1980II}
\begin{eqnarray}
k_{s\rightarrow l}=\frac{\lambda}{\omega_m}\frac{\omega_s}{2\pi}e^{-\beta W}\;,
\end{eqnarray}
where $W=G(r_m)-G(r_s)$ is the barrier height between the small black hole state and the intermediate black hole state on the free energy potential for the RNAdS black holes. So the non-Markovian transition rate for the RNAdS black holes in the intermediate friction regime depends exponentially on the free energy barrier height between the two phases and also on the oscillating frequencies at the initial basin and barrier top as well as the reactive frequency. 

Note that the MFPT is given by the inverse of the transition rate. Therefore, the MFPT in the intermediate friction regime is then given by \begin{eqnarray}\label{intermediate_friction}
\bar{t}=\frac{\omega_m}{\lambda}\frac{2\pi}{\omega_s}e^{\beta W}\;.
\end{eqnarray}

It should be noted that the transition rate for the intermediate friction regime is also valid in the large friction limit. As a consistent check, one can show that the MFPT in the large friction limit Eq.(\ref{MFPT_large_exp}) can be derived from the transition rate for the intermediate friction regime. In the large friction limit, from Eq.(\ref{lambda_eq}), one can see that $\lambda\rightarrow 0$, which in turn gives the approximate solution of $\lambda$ as 
\begin{eqnarray}
\lambda=\frac{\omega_m^2}{\hat{\zeta}(0)}\;.
\end{eqnarray}
Substituting the solution into the transition rate, one can get the MFPT formula in the large friction limit.

\subsection{MFPT: weak friction}

In this subsection, we consider the MFPT of the phase transition process in the weak friction regime. In the case of extremely weak friction, the dynamics is governed by energy diffusion or equivalently action diffusion because the energy change will be slow. In fact, Zwanzig \cite{Zwanzig:1959} derived an equation for the time evolution of the probability distribution function in the energy space with the non-Markovian effect \cite{Grote:1982}. It was later shown that Zwanzig's diffusion equation corresponds to the generalized Langevin equation with the memory kernel represented by the time dependent friction and the random noise characterized by a finite correlation time \cite{Carmeli:1983,Carmeli:1982}. The main assumption used in \cite{Carmeli:1983,Carmeli:1982} to derive the diffusion equation in energy space is that the friction kernel will decay to zero in a finite time scale.

By introducing the action variable 
\begin{eqnarray}
I(E)=\oint p(r) dr\;,\;\;
p(r)=\sqrt{2(E-G(r))}\;,
\end{eqnarray}
the probability distribution in the energy space satisfies the Fokker-Planck equation as \cite{Zwanzig:1959,Grote:1982,Carmeli:1983,Carmeli:1982} 
\begin{eqnarray}
\frac{\partial}{\partial t}\rho(I,t)=
\frac{\partial}{\partial I} \left\{2\pi \epsilon(I)\left[2\pi k_B T \frac{\partial}{\partial I} +\omega(I) \right]\rho(I,t)\right\}\;,
\end{eqnarray}
where the angular frequency $\omega(I)$ at the action $I$ and the diffusion coefficient in the energy space $\epsilon(I)$ are defined as 
\begin{eqnarray}\label{epsilon_def}
\omega(I)&=&2\pi \frac{\partial E}{\partial I}\;,\nonumber\\
\epsilon(I)&=&\frac{1}{\omega^2(I)} \int_0^{\infty} \zeta(t) \langle v(0)v(t)\rangle dt\;.
\end{eqnarray}
Hereby $v(t)$ should be obtained by solving the generalized Langevin equation without the dissipation or the friction kernel $\zeta(t)$ and the noise $\eta(t)$ for constant energy $E(I)$ and $\langle v(0)v(t)\rangle$ corresponds to averaging the initial phase.

From the diffusion equation in energy space, the MFPT to reach the final action $I$ from the initial action $I_0$ can be derived \cite{Grote:1982,Carmeli:1983,Carmeli:1982}
\begin{eqnarray}
\hat{t}=\frac{1}{k_B T} \int_{I_0}^I\left\{
\frac{1}{\epsilon(x)}e^{E(x)/k_B T} \int_0^x 
e^{-E(x')/k_B T} dx'
\right\}dx\;.
\end{eqnarray}
By performing the integral approximately, a compact approximate formula for the MFPT in the weak friction regime can be given as \cite{Grote:1982,Carmeli:1983,Carmeli:1982}
\begin{eqnarray}\label{MFPT_small}
\bar{t}=\frac{k_B T}{\omega_s \epsilon(I_m) \omega(I_m)} e^{\beta W}\;,
\end{eqnarray}
where $I_m$ is the action corresponding to the free energy $G(r_m)$ and $\epsilon(I_m)$ can be computed by using Eq.(\ref{epsilon_def}).

Using the parabola approximation of the generalized Gibbs free energy, one can further simplify the analytical expression (\ref{MFPT_small}) of the MFPT. Firstly, for the deterministic oscillatory movement inside the initial potential well, the action $I$ at the energy $G(r_m)$ has the relation
\begin{eqnarray}
I_m=\frac{2\pi W}{\omega_s}\;.
\end{eqnarray}
For the deterministic oscillation, one has 
\begin{eqnarray}
\langle v(0)v(t) \rangle=\frac{W}{2}\left(e^{i\omega_s t}+e^{-i\omega_s t}\right)\;.
\end{eqnarray}
Therefore, according to Eq.(\ref{epsilon_def}), $\epsilon(I_m)$ can be given by 
\begin{eqnarray}
\epsilon(I_m)=\frac{W}{2\omega^2(I_m)}\mathcal{F}[\zeta(t)](\omega_s)\;, \end{eqnarray}
with $\mathcal{F}[\zeta(t)]$ being the Fourier transform of the friction kernel, which is defined as 
\begin{eqnarray}
\mathcal{F}(\omega)=\int_{-\infty}^{+\infty} e^{i\omega t} \zeta(t)dt\;.
\end{eqnarray}
Finally, the analytical expression of the MFPT in the weak friction regime is given by
\begin{eqnarray}\label{low_friction}
\bar{t}= \frac{2k_B T}{\mathcal{F}[\zeta(t)](\omega_s) W} e^{\beta W}\;.
\end{eqnarray}

\subsection{MFPT: bridging formula}

In the large, intermediate, and weak friction regimes, the MFPTs are amenable to an analytical derivation. We have derived three analytical expressions of the MFPTs in different friction regimes, which are given by Eq.(\ref{MFPT_large_exp}), (\ref{intermediate_friction}), and (\ref{low_friction}). It should be noted that the expression for the MFPT in the large friction regime is included in the expression in the intermediate friction regime. Now we present a bridging formula of the MFPT that is valid in the whole friction regime \cite{Carmeli:1984,Schuller:2019ega,Hanggi:1990}. 

Note that the MFPT is proportional to the strength of the friction kernel in the intermediate-strong friction regime while it is inversely proportional to the friction strength in the weak friction regime. This is to say that $\bar{t}_{low friction}$ given by Eq.(\ref{low_friction}) will become shorter when the friction coefficient increases and $\bar{t}_{intermediate friction}$ given by Eq.(\ref{intermediate_friction}) will become shorter when the friction coefficient decreases, i.e., $\bar{t}_{low friction}$ and $\bar{t}_{intermediate friction}$ have different behaviors with respect to the friction strength. Therefore, a very simple and intuitive approach to give the bridging formula that is valid in the whole friction regime is to interpolate between the already known formulas \cite{Carmeli:1984,Schuller:2019ega} 
\begin{eqnarray}
\bar{t}=\bar{t}_{low friction} +\bar{t}_{intermediate friction}\;.
\end{eqnarray}
When the friction coefficient approaches to $0$, $\bar{t}_{intermediate friction}$ approaches $0$. When the friction coefficient approaches to infinity, $\bar{t}_{low friction}$ approaches $0$. Therefore, the kinetic time $\bar{t}$ given by the above formula is dominated by $\bar{t}_{low friction}$ in the low friction regime and by $\bar{t}_{intermediate friction}$ in the large friction regime. The above formula gives the well behaved interpolation formula for the kinetic time in the whole friction regime. In the following sections, we will use this formula of the MFPT to investigate the non-Markovian dynamics of the Hawking-Page phase transition and the small/large RNAdS black hole phase transition.

The two analytical results of the MFPTs given by Eq.(\ref{low_friction}) and Eq.(\ref{MFPT_large_exp}) imply that the kinetic time will be a decreasing function of friction in the small friction regime and an increasing function of the friction in the large friction regime. Therefore, in the intermediate friction regime, the kinetic turnover will be an universal property for any kinds of friction kernels.

\section{Effects of time dependent frictions on the dynamics of the black hole phase transition}\label{Num_results}

In this section, we discuss the numerical results of the dynamics of the black hole phase transition represented by the MFPTs from one phase to another phase. As mentioned previously, we mainly focus on two types of phase transition, i.e., Hawking-Page phase transition and the small/large RNAdS black hole phase transition. In the following, we will discuss three types of the friction kernels, i.e., the delta friction, exponentially decayed friction, and oscillatory friction. We will study the dependency of the kinetic time on the ensemble temperature, the friction strength, decay coefficient, and oscillating frequency.

\subsection{Delta friction}

Firstly, we consider the friction of the delta function
\begin{eqnarray}
\zeta(t)=\zeta \delta(t)\;. 
\end{eqnarray}
As mentioned, under this assumption, the non-Markovian Langevin equation reduces to the Markovian case. This assumption gives the Laplace transformation as
\begin{eqnarray}
\hat{\zeta}(\lambda)=\zeta\;,
\end{eqnarray}
and 
\begin{eqnarray}
\frac{\lambda}{\omega_m}=\left(\sqrt{\frac{\zeta^2}{4\omega_m^2}+1}-\frac{\zeta}{2\omega_m}\right)\;.
\end{eqnarray}
Then the transition rate from one stable state to another stable state for the delta friction is given by 
\begin{eqnarray}\label{delta_rateeq_large}
k_{s\rightarrow l}=\left(\sqrt{1+\frac{\zeta^2}{4\omega_m^2}}-\frac{\zeta}{2\omega_m}\right)\frac{\omega_s}{2\pi}e^{-\beta W}\;.
\end{eqnarray}
Note that this expression is only valid in the intermediate-large friction regime. For the small friction regime, we have 
\begin{eqnarray}\label{delta_rateeq_small}
\bar{t}= \frac{k_B T}{\zeta W} e^{\beta W}\;.
\end{eqnarray}
This is consistent with the the result that derived in \cite{Li:2021vdp}. There is a subtlety in deriving the above result when considering the boundary condition in deriving the MFPT for the weak friction. For small friction case, we explore the diffusion in the energy space. The first passage time is counted by the time that the initial state reaches the potential energy of the intermediate black hole state at the barrier top of the free energy landscape. Therefore, a factor of $1/2$ should be taken into account in the expression of the MFPT in the weak friction regime.

The two results given by (\ref{delta_rateeq_large}) and (\ref{delta_rateeq_small}) implies that the kinetic time will be a decreasing function of friction in the small friction regime and an increasing function of the friction in the large friction regime. Therefore, in the intermediate friction regime, there will be kinetic turnover. This reflects the fact that at high friction, the diffusion is along the order parameter $r$ and increasing friction leads to slower kinetics while at low friction, the diffusion dynamics is along the energy. Increasing the friction leads to larger perturbations to the original conserved system and this gives rise to faster the diffusion kinetics in energy space.

\subsubsection{Dynamics of Hawking-Page phase transition}

For the Hawking-Page phase transition, this expression can be used to calculate the MFPT of the phase transition process from the large SAdS black hole state to the thermal AdS state by replacing the $\omega_m$ with $\omega_s$ and $\omega_s$ with $\omega_l$. Because there is not a well-defined parabolic approximation at the well of the thermal AdS state on the free energy landscape, the MFPT from the thermal AdS state to the large SAdS black hole state cannot be computed by using the analytical expression above. Therefore, we focus on the kinetic time of Hawking-Page phase transition from the large SAdS black hole state to the thermal AdS state.

\begin{figure}
  \centering
  \includegraphics[width=6cm]{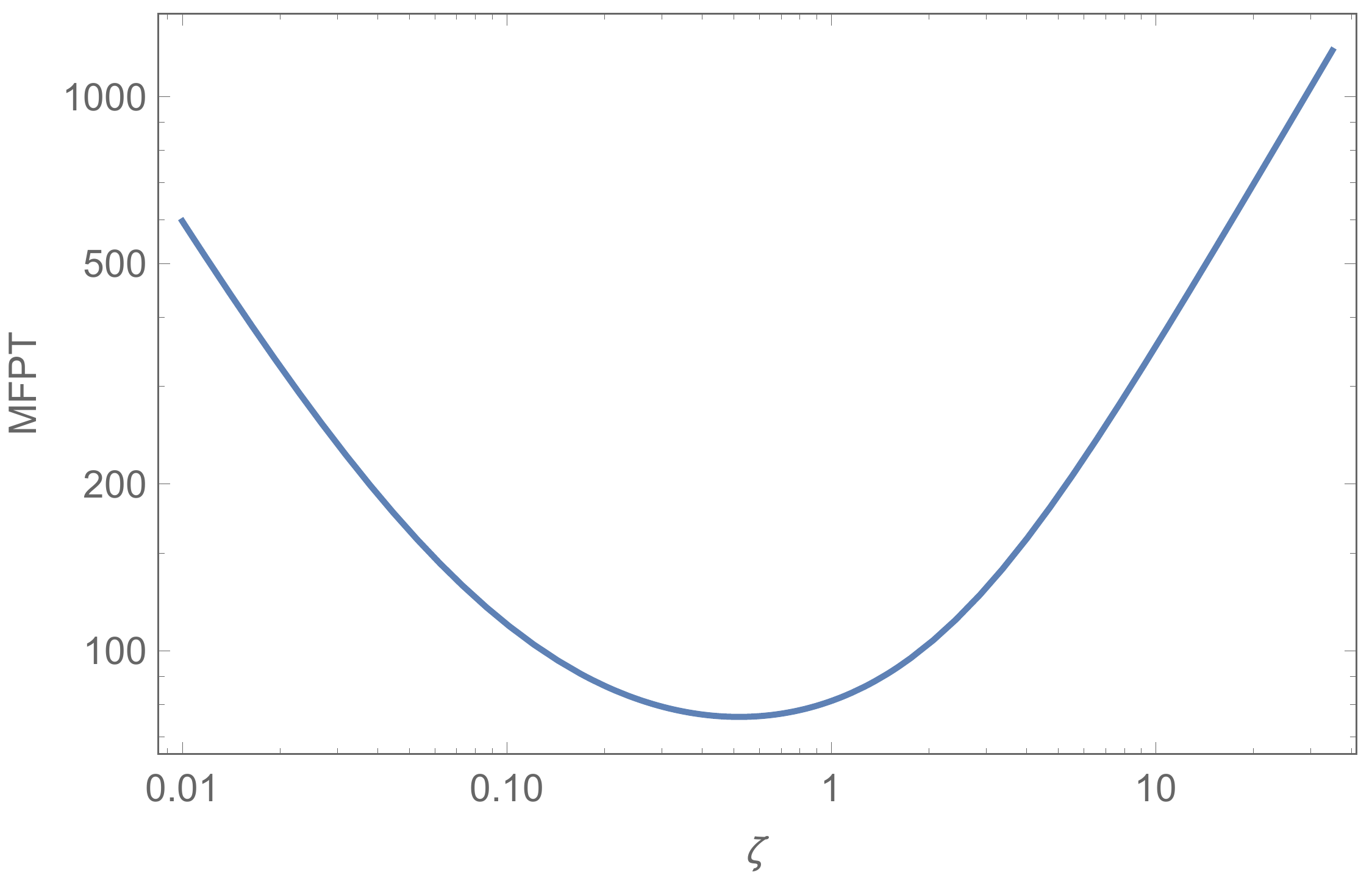}\\
  \caption{The dependency of the MFPT of the Hawking page phase transition from the large SAdS black hole state to the thermal AdS state on the friction coefficient for the delta friction. In the plot, $L=1$, and $T=0.5$. 
  }\label{MFPTvsZeta_delta_HP}
\end{figure}

In Figure \ref{MFPTvsZeta_delta_HP}, we plot the dependency of the kinetic time of Hawking-Page phase transition from the large SAdS black hole state to the thermal AdS state on the friction for the delta-type friction kernel. There is a kinetic turnover around $\zeta=1$. This kind of phenomenon has been observed in \cite{Li:2021vdp} where the kinetic turnover for the RNAdS black hole phase transition is considered. The behavior of the kinetic time in the low and the high friction can be interpreted as follows. In the overdamped regime, the friction is large. Then the dynamics of the black hole phase transition is diffusive in the order parameter space (horizon size). Thus, the larger friction will impede the motions along the order parameter. As a result, the kinetics will be slowed down. On the other hand, in the under damped regime, the friction is small. Then, the dynamics of the of the black hole phase transition is diffusive in the energy space. This is because, without the friction, the energy is conserved. Adding a bit of the friction promotes the drift or diffusive dynamics in energy space. Therefore, the resulting kinetics will be faster when the friction is increased. This explains the non-monotonic behaviors of the kinetics of the black hole phase transition. This indicates that when the coupling strength among the black hole degrees of freedom is strong, the black hole phase transition kinetics is slower when the coupling is increased. On the other hand, when the coupling strength among the black hole degrees of freedom is weak, the black hole phase transition kinetics is faster when the coupling is increased.

\begin{figure}
  \centering
  \includegraphics[width=6cm]{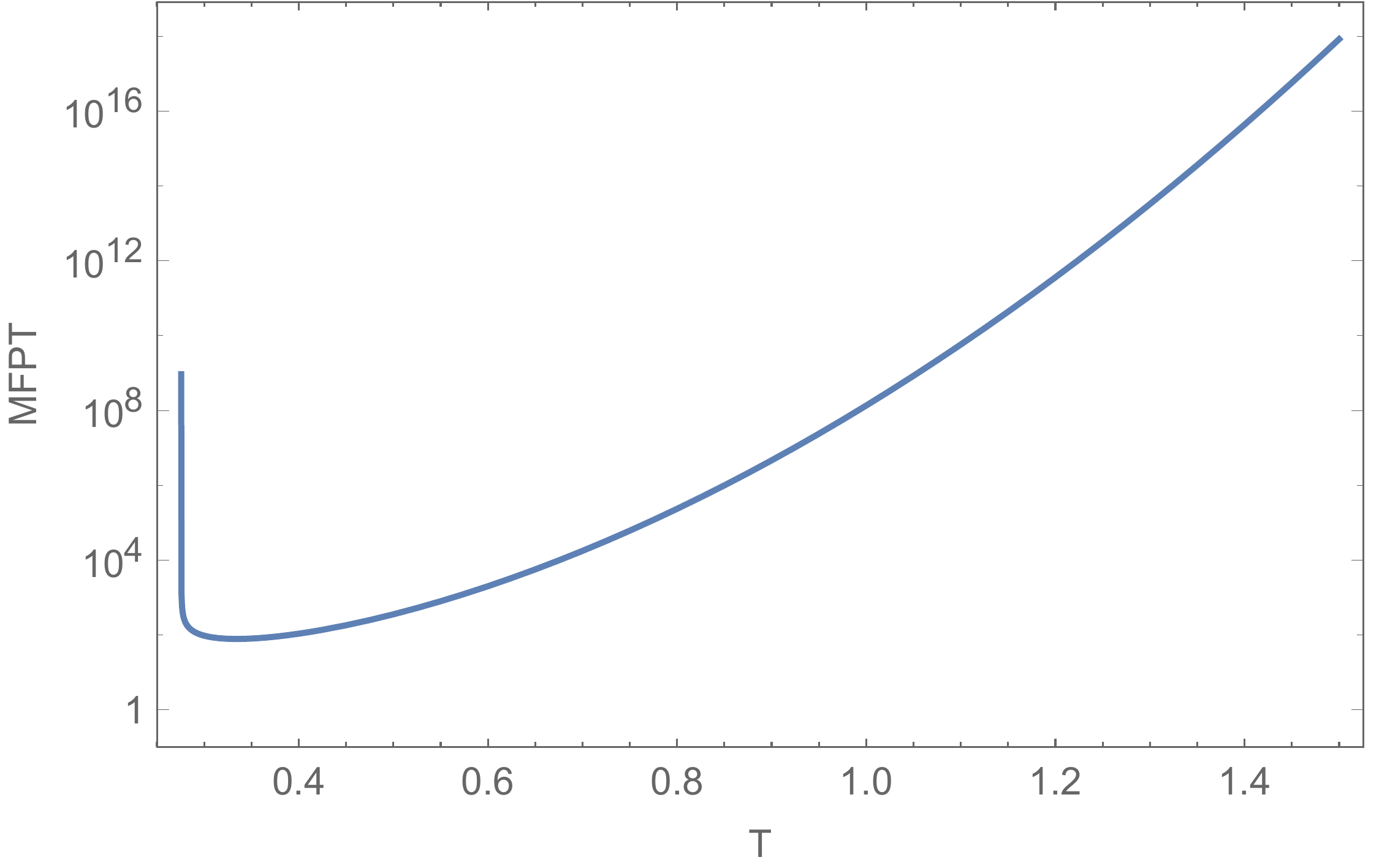}\\
  \caption{The dependency of the MFPT of the Hawking page phase transition from the large SAdS black hole state to the thermal AdS state on the friction coefficient for the delta friction. In the plot, $L=1$, and $\zeta=10$. 
  }\label{MFPTvsT_delta_HP}
\end{figure}

In Figure \ref{MFPTvsT_delta_HP}, we plot the MFPT as a function of the ensemble temperature for the delta friction. For this plot, $\zeta$ is selected to be $10$. According to the previous discussion, this $\zeta$ is in the range of large friction regime. We have tried several different $\zeta$ in different regime. The same behavior can be observed. When $T>0.4$, the MFPT is the increasing function of the ensemble temperature. The reason behind is that the potential barrier height monotonically increases with the temperature. When $T>0.4$, the potential barrier is relatively larger than $k_B T$. In this case, the kinetic time is mainly determined by the factor $e^{\beta W}$ giving a combined effect at temperature and barrier. When $T\rightarrow T_{min}$, the MFPT is divergent. This is caused by $W\rightarrow 0$ in Eq.(\ref{delta_rateeq_small}) and $\omega_l\rightarrow 0$ in Eq.(\ref{delta_rateeq_large}) when $T\rightarrow T_{min}$. This reflects that the wrong application of the analytical results, because the analytical results are valid only under the condition of $W/k_B T\ll 1$. In principle, the MFPT should approach zero when the potential barrier approaches zero. The accurate results should invoke the numerical computation of the MFPT by solving the generalized Langevin equation. This is out of the scope of the present work.

\subsubsection{Dynamics of RNAdS black hole phase transition}

For the RNAdS black hole phase transition, the similar expression for the transition rate or the MFPT from the large black hole state to the small black hole state can be obtained by replacing $W$ as the barrier height between the large black hole state and the intermediate black hole state. This result coincides with that obtained from the Markovian dynamics. However, one should note that this analytical result of the transition rate is only valid when the barrier height between the small black hole state and the intermediate black hole state is much bigger than $k_B T$.

\begin{figure}
  \centering
  \includegraphics[width=6cm]{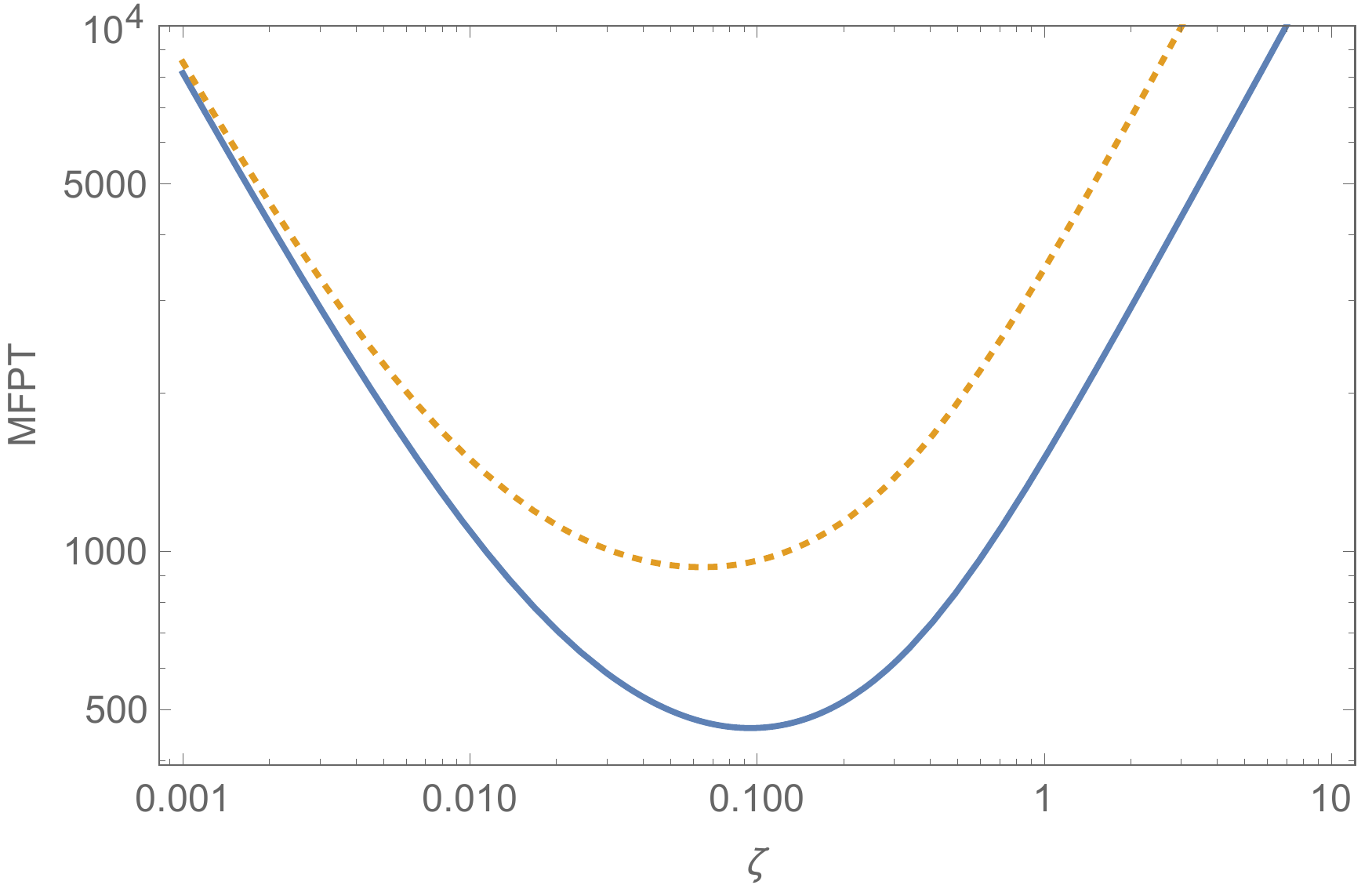}\\
  \caption{MFPT as the function of friction for the small/large RNAdS black hole phase transition. In the plot, $Q=1$, $P=0.4 P_c$, and $T=0.0298$. The temperature is selected to be the phase transition temperature, which on the free energy landscape, the depths of the left and right wells are equal. The solid line is for the transition from the small black hole state to the large, and the dotted line is for the inverse process.   
  }\label{MFPTvsZeta_delta_RNAdS}
\end{figure}

In Figure \ref{MFPTvsZeta_delta_RNAdS}, we plot the MFPTs of the small/large RNAdS black hole phase transition for the delta friction. The temperature is selected to be the phase transition temperature, where on the free energy landscape, the depths of the left and right wells are equal. At this specific temperature, the ratio of the barrier height $W$ and $k_B T$ is $W/k_B T=3.23$, which is larger than unity. The condition that the analytical results applies is satisfied. In this plot, we consider the two different phase transition processes, i.e., the process from the small black hole to the large black hole and its inverse process. The numerical results are plotted in solid line and dotted line respectively. As expected, there are turnover point in the kinetics of the two processes. The reason has been explained in the last section. Another observation is that when $\zeta$ is small, the MFPTs for the two processes are equal, while for large $\zeta$, the MFPTs for the two processes have the same slopes. The reason is that Eq.(\ref{delta_rateeq_small}), which is dominant in the small $\zeta$ regime, is independent of the shape of the free energy landscape and only determined by the barrier height. When the temperature is selected to be the phase transition temperature, the barrier heights for the two phase transition processes are equal. Therefore, in the small $\zeta$ regime, the MFPTs are the same. In the large friction regime, the MFPTs for the two phase transition processes differ by the prefactor $\omega_s$ or $\omega_l$. Therefore, their dependencies on the friction are the same and the MFPTs only differ a constant factor determined by the ratio of $\omega_s/\omega_l$.

\begin{figure}
  \centering
  \includegraphics[width=6cm]{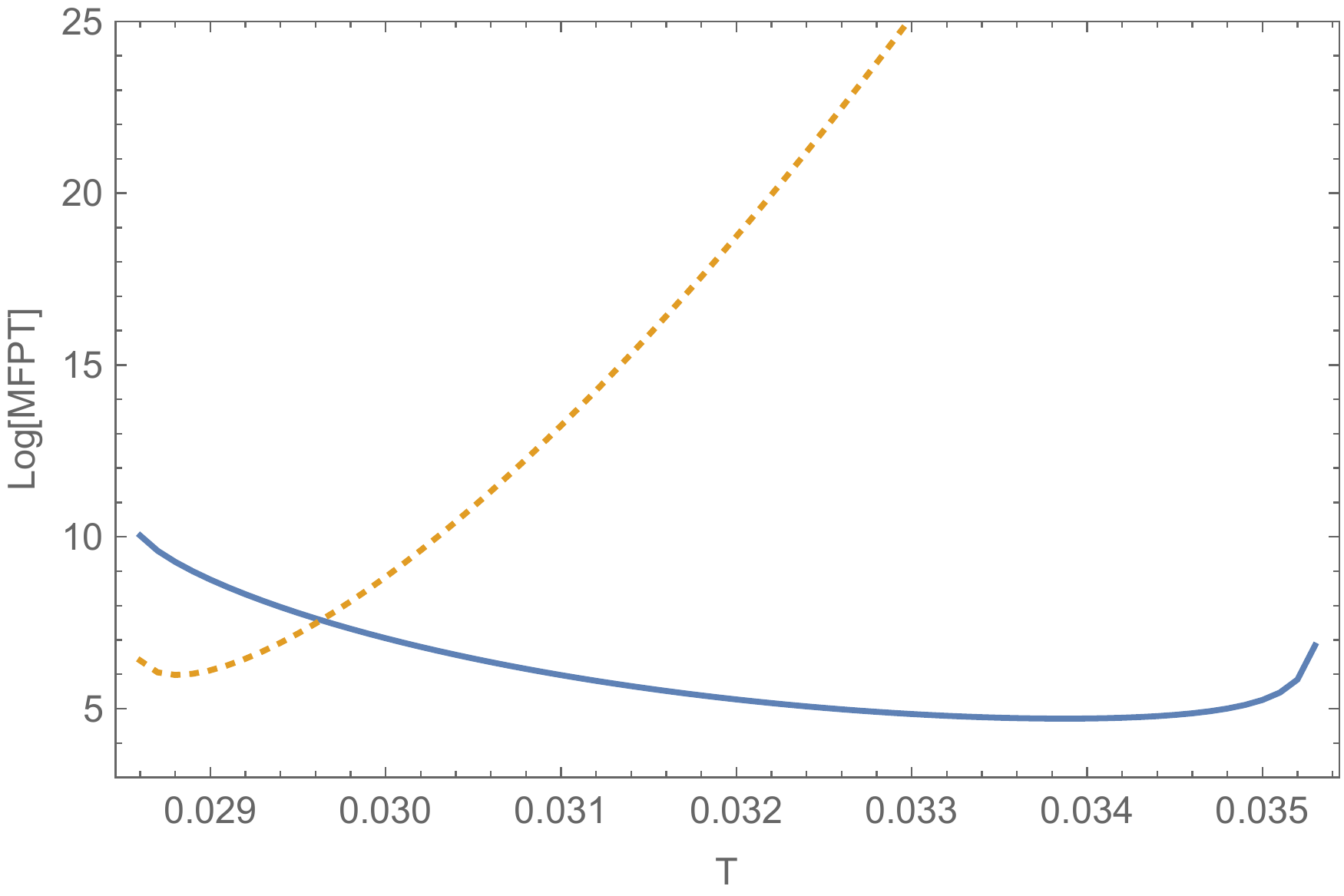}\\
  \caption{MFPT of the RNAdS black hole phase transition as the function of the ensemble temperature for the delta friction. In the plot, $P=0.4 P_c$, $Q=1$, and $\zeta=1$. The solid line is for the transition process from the small black hole to the large, and the dotted line is for the inverse process. }\label{MFPTvsT_delta_RNAdS}
\end{figure}

In Figure \ref{MFPTvsT_delta_RNAdS}, we plot the MFPT of the RNAdS black hole phase transition as a function of the ensemble temperature for the delta friction. In this plot, $\zeta$ is selected to be $1$. According to the discussion of Figure \ref{MFPTvsZeta_delta_RNAdS}, this $\zeta$ is also in the range of large friction regime. The plots for different values of $\zeta$ in different damping regime show the same behavior. Therefore, we have plotted the case of $\zeta=1$ as a  representative example. The solid line is for the transition process from the small black hole to the large one, and the dotted line is for the inverse process. As noted previously, the analytical results are only valid when $W/k_B T\ll 1$. Therefore, the solid line's plot of the MFPT at the high temperature and the dotted line's plot of the MFPT at the low temperature do not reflect the real behavior because of the low potential barrier on the free energy landscape. In the following discussion, we will ignore theses parts in the plots. For the phase transition from the small black hole to the large, it is shown that the MFPT is a decreasing function of the ensemble temperature. On the other hand, for the phase transition process from the large black hole to the small one, the MFPT is the increasing function of the ensemble temperature. This behavior is mainly caused by the behavior of the barrier heights on the free energy landscape. When increasing the temperature, the barrier height between the small black hole and the intermediate black hole decreases giving rise to faster kinetic phase transition rate while the barrier height between the large black hole and the intermediate black hole increases giving rise to slower kinetic phase transition rate. If the friction is fixed, the kinetics of the phase transition is mainly determined by the potential barrier depth on the free energy landscape.

\subsection{Exponentially decayed friction}

For the exponentially decayed friction, we mainly focus on the non-Markovian effects on kinetics of the black hole phase transition. In other words, we consider the effects of the time scale of the friction kernel. The time scale of the friction kernel describes the correlation time of the effective thermal bath of the underlying microscopic degrees of freedom for the black hole. In order to compare with the delta friction, i.e. Markovian dynamics, we consider the exponentially decayed friction which is given by 
\begin{eqnarray}
\zeta(t) = \frac{\zeta}{\gamma} e^{-\frac{|t|}{\gamma}}\;,
\end{eqnarray}
where $\zeta$ represents the strength of the friction and $\gamma$ is the time scale of the friction kernel. When $\gamma\rightarrow 0$, the exponentially decayed friction is reduced to the delta friction discussed in the last subsection. 

For the exponentially decayed friction, the Laplace transformation is given by \begin{eqnarray}\label{zeta_laplace}
\hat{\zeta}(\lambda)=\frac{\zeta}{1+\lambda\gamma}\;.
\end{eqnarray}
The equation that determines the prefactor $\lambda$ of the transition rate becomes a cubic equation from Eq.(\ref{lambda_eq}) 
\begin{eqnarray}
\gamma\lambda^3+\lambda^2+\left(\zeta-\gamma\omega_m^2\right)\lambda-\omega_m^2=0\;.
\end{eqnarray}
The $\lambda$ is given by the largest root of the above equation. By substituting the root into the analytical expression, one can calculate the MFPT of the phase transition process in the intermediate-strong friction regime.

For the small friction case, one should calculate the Fourier transformation of the friction kernel, which is given by 
\begin{eqnarray}
\mathcal{F}(\omega)=\frac{2\zeta}{1+\omega^2\gamma^2}
\end{eqnarray}
When $\gamma\rightarrow 0$, it gives $\mathcal{F}(\omega)=2\zeta$, which in consequence gives the MFPT in the weak friction regime for the delta friction. By taking the value of $\omega$ as the oscillating frequency of the initial potential well on the free energy landscape, one can get the MFPT of the transition process in the weak friction regime. It can be shown that $\mathcal{F}(\omega)$ is proportional to the friction coefficient $\zeta$, which implies that the MFPT in the weak friction regime is also inversely proportional to the strength coefficient of the friction kernel. Therefore, it is anticipated that there will also be a turnover point in the kinetics of the exponentially decayed friction when varying the friction coefficient.

\subsubsection{Dynamics of Hawking-Page phase transition}

For the Hawking-Page phase transition, we calculate the kinetic time from the large SAdS black hole state to the thermal AdS state. As mentioned, the analytical expressions for the MFPT should be properly modified, because these expressions are derived by assuming that the initial state is in the left well of the free energy landscape and the initial state for Hawking-Page phase transition is in the right well.

\begin{figure}
  \centering
  \includegraphics[width=6cm]{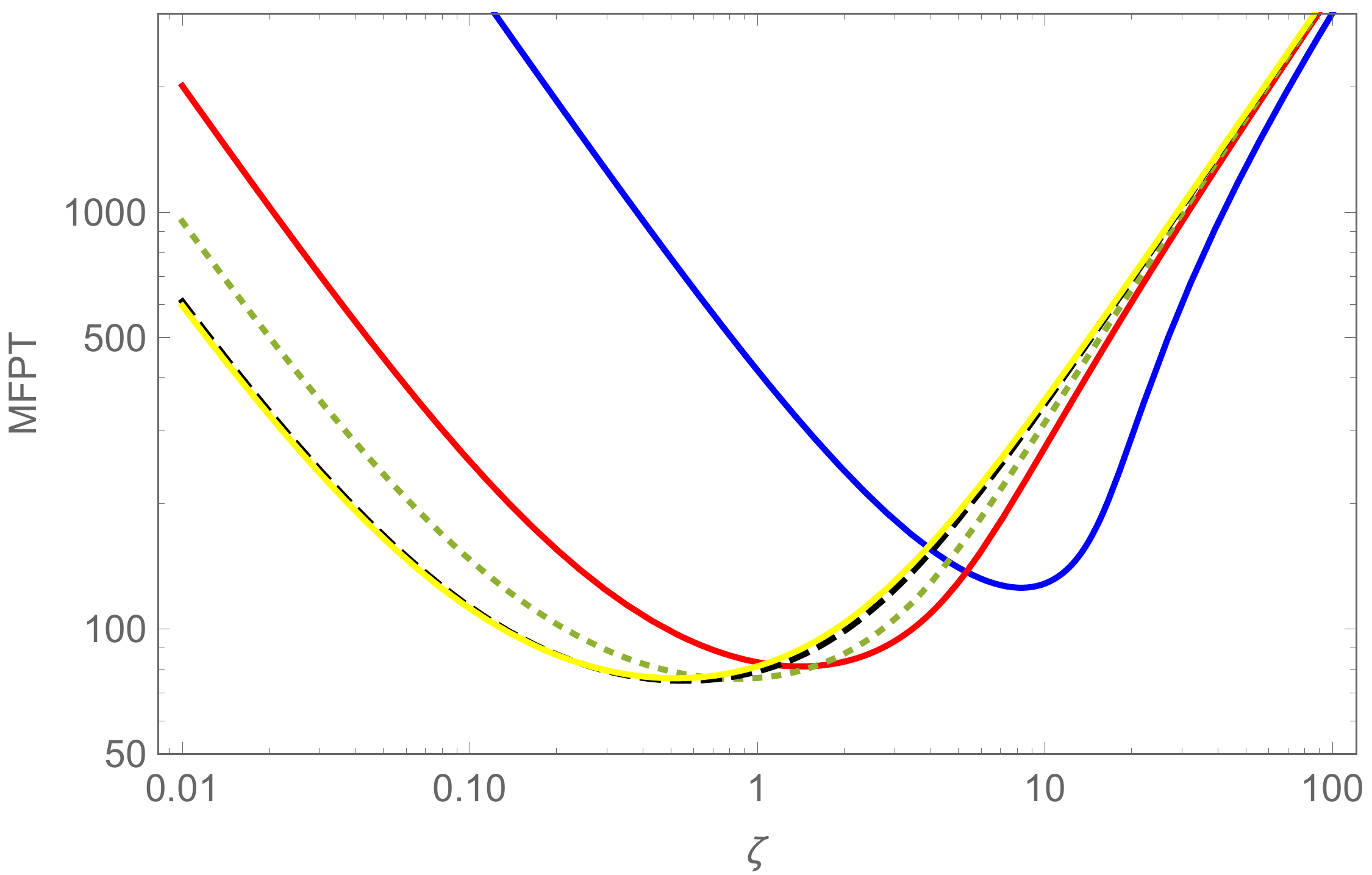}\\
  \caption{The dependency of the MFPT of the Hawking-Page phase transition from the large SAdS black hole state to the thermal AdS state on the friction coefficient for the exponentially decayed friction at different decay coefficients. In the plot, $L=1$, and $T=0.5$. The blue, red, dotted, dashed lines are for $\gamma=5, 1, 0.5, 0.1$. For the convenience of comparing, we also plot the delta friction friction's result in yellow.  
  }\label{MFPTvsZeta_decay_HP}
\end{figure}

In Figure \ref{MFPTvsZeta_decay_HP}, we plot the dependency of the MFPT of the Hawking-Page phase transition from the large SAdS black hole state to the thermal AdS state on the friction coefficient $\zeta$ for the exponentially decayed friction at different decay coefficient $\gamma$. When taking the decay coefficient to zero, the exponentially decayed friction will be reduced to the delta friction friction. Their qualitatively similar behavior of the kinetic times is anticipated. It is shown that there are turnover points in the kinetics with respect to friction strength for different decay coefficients. This implies that the kinetic turnover is an universal property not only for the Markovian dynamics but also for the non-Markovian dynamics. The numerical results verify the general observation obtained from the analytical expressions of the MFPT in the small and large friction regimes, i.e., the MFPT in the weak friction regime is inversely proportional to the strength coefficient of the friction kernel while in the strong friction regime, it is proportional to the friction coefficient. Therefore, in the intermediate friction regime, there will be kinetic turnover.

The plot for small $\gamma$ in Figure \ref{MFPTvsZeta_decay_HP} should approximate the result of the Markovian dynamics while other plots reflect the effect of the non-Markovian dynamics on the kinetics of the Hawking-Page phase transition. It can be seen that the plot of $\gamma=0.1$ is very close to the delta friction's plot, which is plotted in yellow for the convenience of comparing. By comparing the plots of $\gamma=0.5, 1, 5$ with the plot of the Markovian result, we can conclude the non-Markovian effects on the kinetics. For the small $\zeta$ (for example $\zeta<1$), the phase transition process of the non-Markovian model is slower than the Markovian dynamics. Increasing the decay coefficient or time scale $\gamma$ will slow down the phase transition process. In the strong friction regime ($\zeta>10$), the dynamics of the non-Markovian model is faster than the dynamics of the Markovian model. Increasing the decay coefficient or time scale $\gamma$ will speed up the phase transition process. In the intermediate friction regime, the dynamics of the phase transition depends on the friction strength and the decay coefficient. No general conclusion can be obtained in the intermediate friction regime. 

\begin{figure}
  \centering
  \includegraphics[width=6cm]{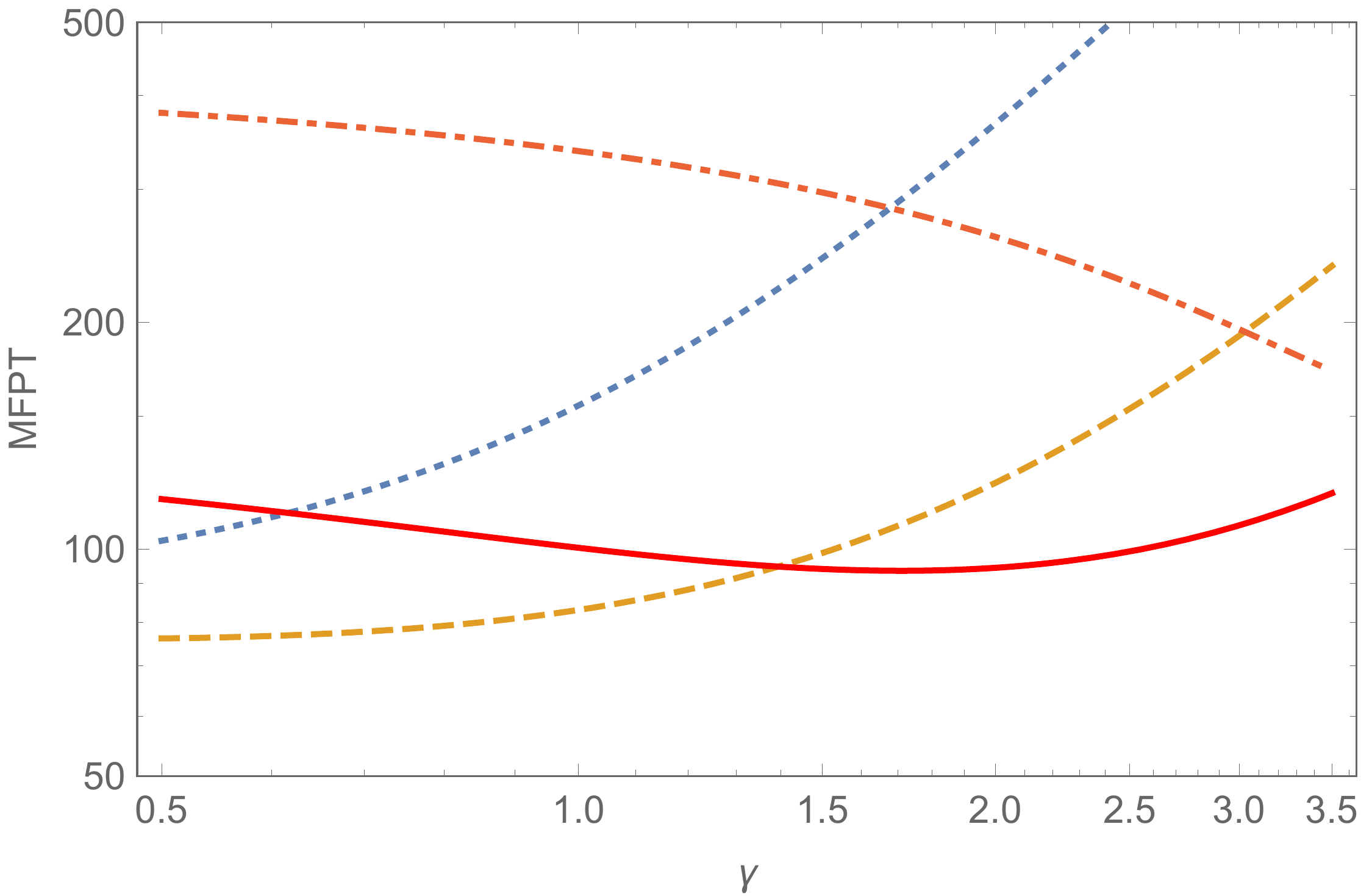}\\
  \caption{The dependency of the MFPT of the Hawking-Page phase transition from the large SAdS black hole state to the thermal AdS state on the decay coefficient for the exponentially decayed friction. In the plot, $L=1$, and $T=0.5$. The dotted, dashed, red, dotdashed lines are for $\zeta=0.2, 1, 3.5, 12$.  
  }\label{MFPTvsGamma_decay_HP}
\end{figure}

In Figure \ref{MFPTvsGamma_decay_HP}, the dependency of the MFPT of the Hawking-Page phase transition on the decay coefficient or the time scale for different friction strength are plotted. For weak friction ($\zeta=0.2, 1$), the MFPT is the increasing function of $\gamma$, while for strong friction ($\zeta=12$), the MFPT is the decreasing function of $\gamma$. The reason behind can be interpreted from the analytical expressions. From Eq.(\ref{lambda_eq}), (\ref{intermediate_friction}) and (\ref{zeta_laplace}), one can deduce that in the large friction regime, $\bar{t}\sim(\zeta-\gamma\omega_s^2)$. Note that $\omega_m$ should be replaced by $\omega_s$ because the intermediate state for the Hawking-Page phase transition is the small SAdS black hole state. In the small friction regime, $\bar{t}\sim 1/\mathcal{F}(\omega_l)\sim(1+\omega_l^2 \gamma^2)$. Note that the initial state is set to be the large SAdS black hole. For the weak friction, the MFPT is proportional to the decay coefficient or the time scale while for the strong friction, the MFPT is negatively correlated to the decay coefficient or time scale. This is to say that the non-Markovian effect speeds up the transition process in the strong friction regime and slows down the process in the weak friction regime. For the intermediate friction $\zeta=3.5$, the competition of the two behaviors leads to the non-monotonic behavior of the transition rate with respect to the decay coefficient $\gamma$ as shown in Figure \ref{MFPTvsGamma_decay_HP}. Note that the speed up behavior of the transition process in the large friction regime has been observed previously \cite{Li:2021}. The reason behind is that the non-Markovian effect introduces more time scales and therefore equivalently more degrees of freedom. The speed up behavior can be explained by the possible "short-cut" path in the equivalent multi-order parameters space. In the weak friction regime, the slow down behavior of the transition dynamics is due to the adiabatic energy transfer effects \cite{Grote:1982}.

\begin{figure}
  \centering
  \includegraphics[width=6cm]{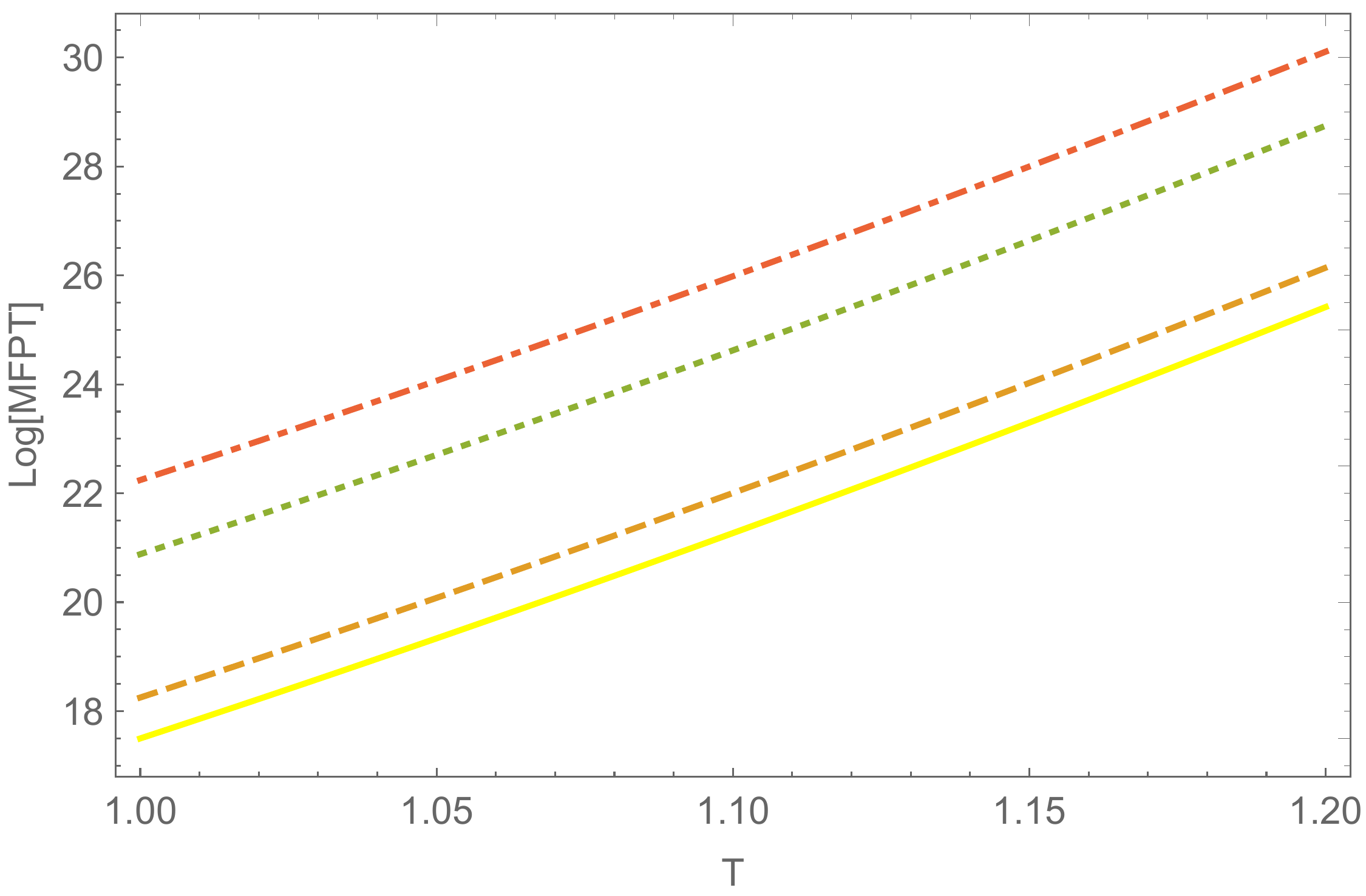}
  \includegraphics[width=6cm]{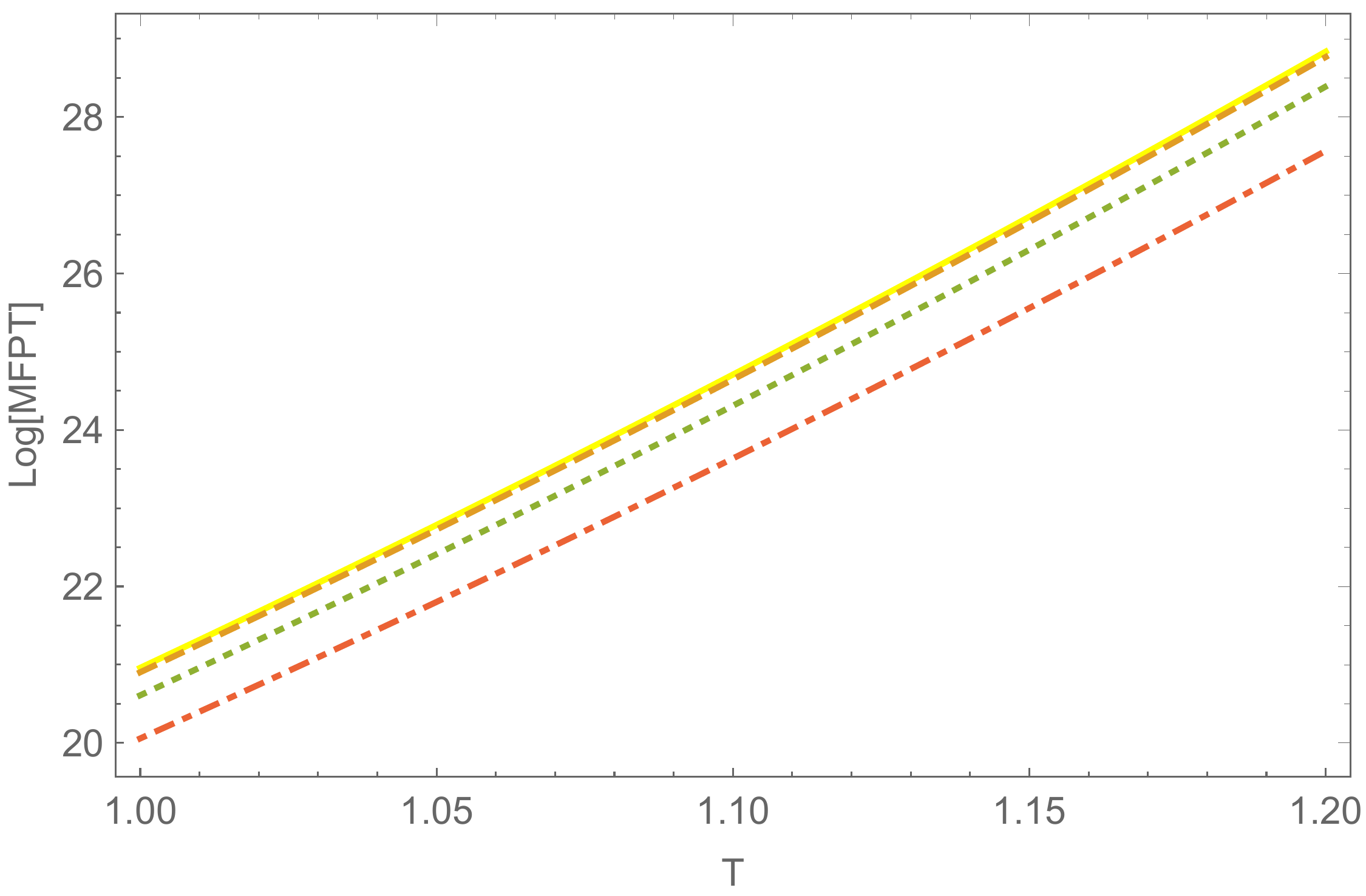}\\
  \caption{The MFPT of the Hawking-Page phase transition from the large SAdS black hole state to the thermal AdS state as a function of the ensemble temperature for the exponentially decayed friction at different decay coefficients. In the left plot, $L=1$, and $\zeta=0.1$. In the right plot, $L=1$, and $\zeta=100$. The yellow, dashed, dotted, and dotdashed lines are for $\gamma=0.1, 1, 5, 10$. 
  }\label{MFPTvsT_decay_HP}
\end{figure}

In Figure \ref{MFPTvsT_decay_HP}, we plot the dependency of the MFPT of the Hawking-Page phase transition from the large SAdS black hole state to the thermal AdS state on the ensemble temperature $T$ for the exponentially decayed friction at different friction coefficient $\zeta$ and decay coefficient $\gamma$. As explained previously, the low temperature behavior of the MFPT is not exact, because the barrier height on the free energy landscape at the low temperature approaches zero. We only consider the relatively high barrier case. The MFPT is shown to be an increasing function of the ensemble temperature $T$, which is similar to the delta friction case. The reason behind is that when the ensemble temperature increases, the barrier height between the large SAdS black hole state and the small SAdS black hole state increases too. The high barrier will slow down the kinetics of the Hawking-Page phase transition. For the different friction coefficient $\zeta$ or the different decay coefficient $\gamma$, we can observe similar behaviors. The yellow plot for $\gamma=0.1$ can be regarded as the result for the Markovian dynamics. The left panel is plotted for small friction while the right panel is for the large friction. Comparing the two figures, we can conclude that in the weak friction regime, the non-Markovian effect slows down the phase transition dynamics while in the strong friction regime, the non-Markovian effect speeds up the transition process. The reason has been explained in the last paragraph.

\subsubsection{Dynamics of RNAdS black hole phase transition}

For the RNAdS black holes, we calculate the kinetic times of the phase transition process from the small black hole to the large black hole and its inverse process.

\begin{figure}
  \centering
  \includegraphics[width=6cm]{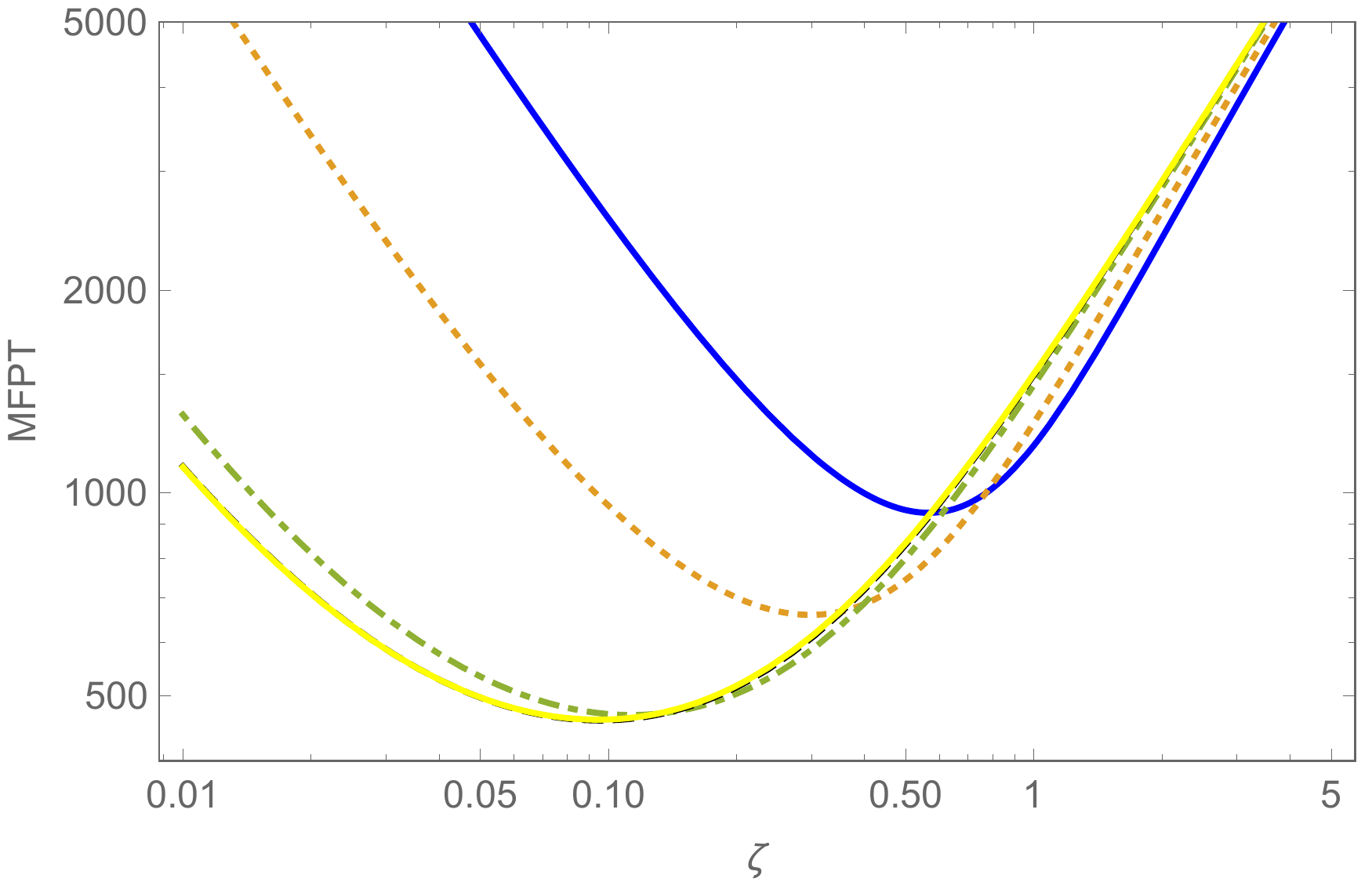}
  \includegraphics[width=6cm]{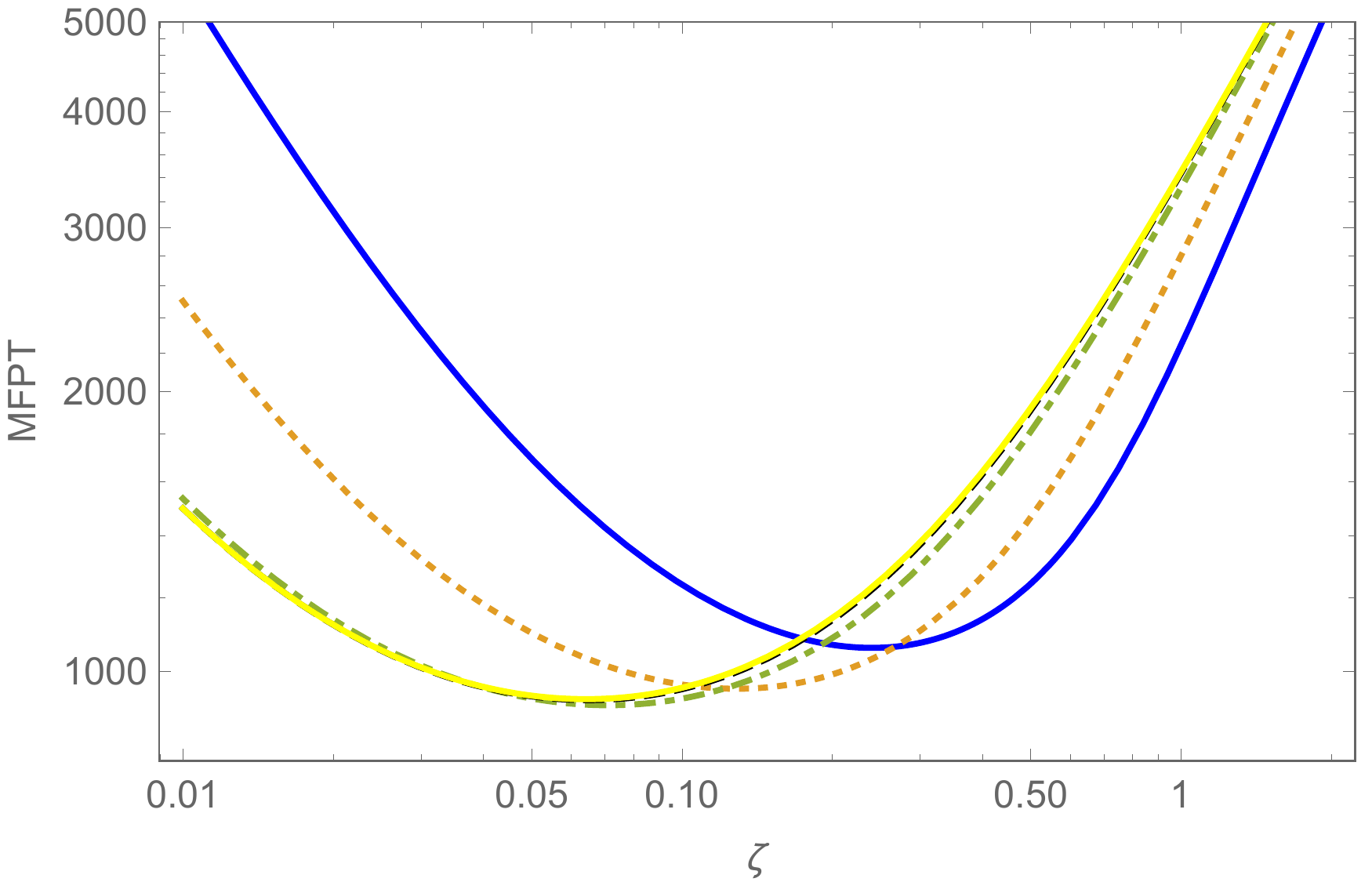}\\
  \caption{The dependency of the MFPT of the RNAdS black hole phase transition on the friction coefficient for the exponentially decayed friction. The left panel is for the transition process from the small black hole state to the large black hole state and the right panel is for the inverse process. In the plots, $Q=1$, $P=0.4 P_c$, and $T=0.0298$. The temperature is selected to be the phase transition temperature, which on the free energy landscape, the depths of the left and right wells are equal. The blue, dotted, dotdashed, and dashed lines are for $\gamma=10, 5, 1, 0.1$. The yellow line is the plot for the delta friction.  }\label{MFPTvsZeta_decay_RNAdS}
\end{figure}

In Figure \ref{MFPTvsZeta_decay_RNAdS}, we plot the MFPTs of the RNAdS black hole phase transition as the function of the friction coefficient $\zeta$ at different decay coefficient or time scale $\gamma$. The upper panel is for the transition process from the small black hole to the large black hole, and the lower one is for the inverse process. All the plots show that there are turnover points in the kinetics of the transition processes, which indicate that the kinetic turnover is an universal property in the black hole transition dynamics. From the two plots, we can also conclude that for the RNAdS black hole phase transition, in the weak friction regime, the non-Markovian effect slows down the phase transition dynamics while in the strong friction regime, the non-Markovian effect speeds up the transition process. This conclusion is also universal for the black hole phase transition. The reason behind has been interpreted in the last subsection.

\begin{figure}
  \centering
  \includegraphics[width=6cm]{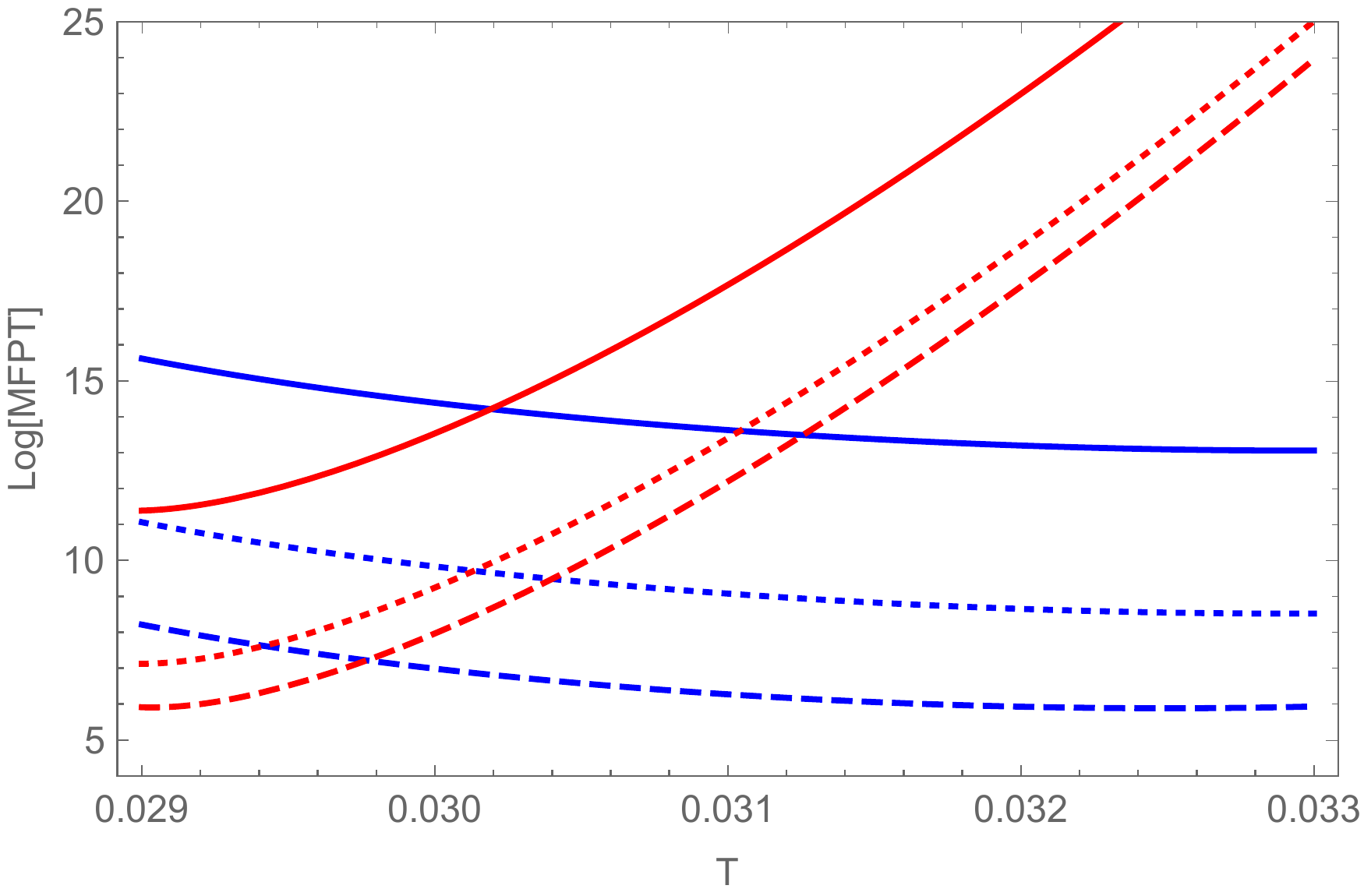}
  \includegraphics[width=6cm]{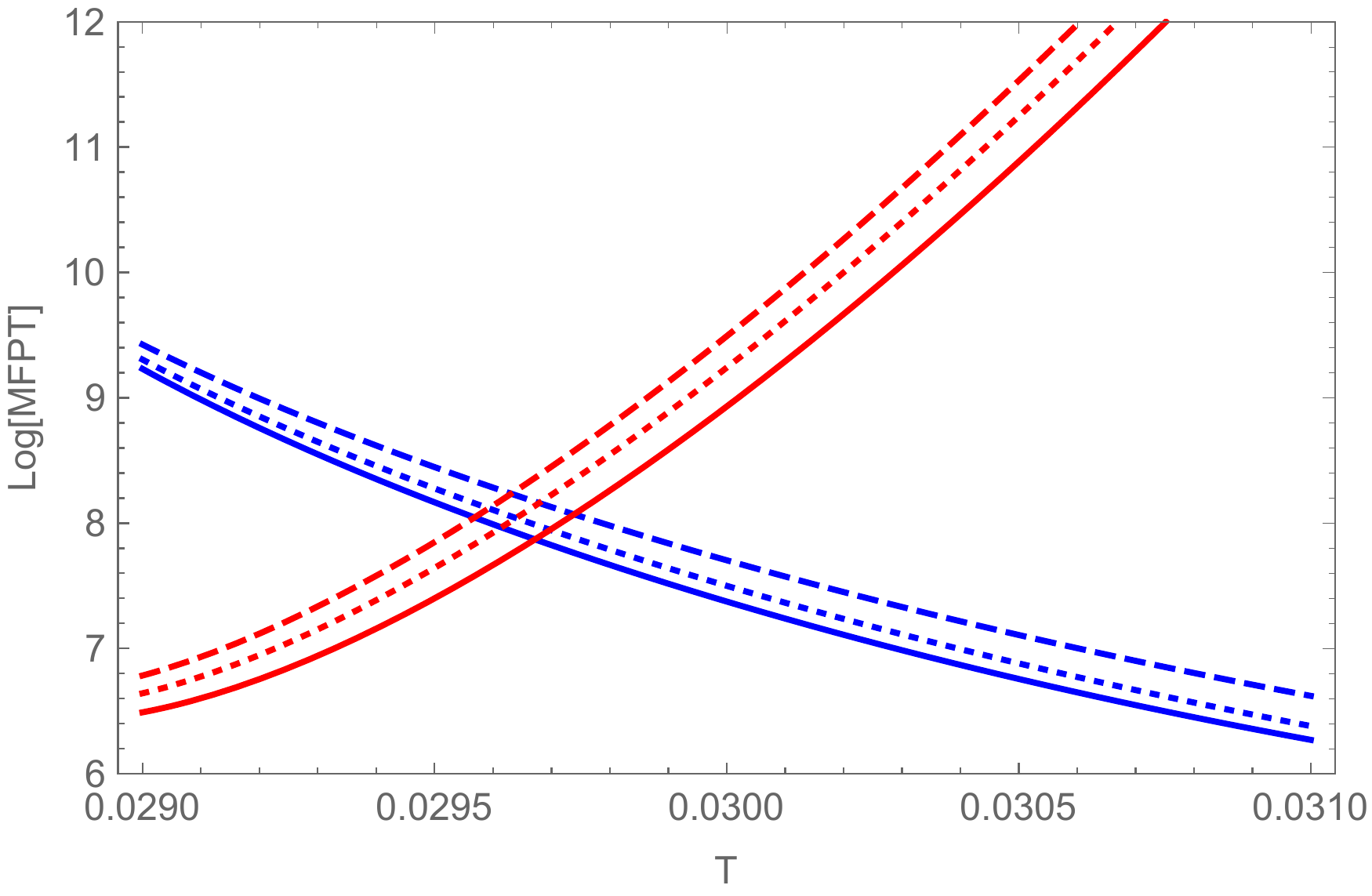}\\
  \caption{The dependency of the MFPT of the RNAdS black hole phase transition for the exponentially decayed friction at weak friction regime. In the plot, $Q=1$, and $P=0.4 P_c$. In the left panel, $\zeta=0.01$ and the solid, dotted, and dashed lines are for $\gamma=100, 10, 1$. In the right panel, $\zeta=2$ and the solid, dotted, and dashed lines are for $\gamma=20, 10, 0.1$. The blue lines are the plots for the processes from the small black hole state to the large black hole state while the red lines are for the inverse processes. 
  }\label{MFPTvsT_decay_RNAdS}
\end{figure}

In Figure \ref{MFPTvsT_decay_RNAdS}, we plot the MFPTs of the RNAdS black hole phase transition as the function of the ensemble temperature. The left panel is plotted in the weak friction regime, and the right panel is plotted in the strong friction regime. From the plots, we can conclude that the transition process from the small black hole state to the large black hole state becomes easier and the inverse process becomes harder when increasing the ensemble temperature. The reason is that the dynamics is dominated by the barrier height that the initial black hole state needed to pass through in the transition process. Another observation is the effect of the non-Markovian dynamics on the black hole phase transition. The previous conclusion can be explicitly shown in the two plots for different decay coefficients or time scales. From the two plots, it can be concluded that for the RNAdS black hole phase transition, in the weak friction regime, the non-Markovian effect slows down the phase transition dynamics while in the strong friction regime, the non-Markovian effect speeds up the transition process.

\subsection{Oscillatory friction}

In this subsection, we consider the final model of the time dependent friction. The friction kernel is oscillatory in time. For the oscillatory friction, we mainly consider the effects of the oscillating frequency on the kinetics of the black hole phase transition.

\begin{figure}
  \centering
  \includegraphics[width=6cm]{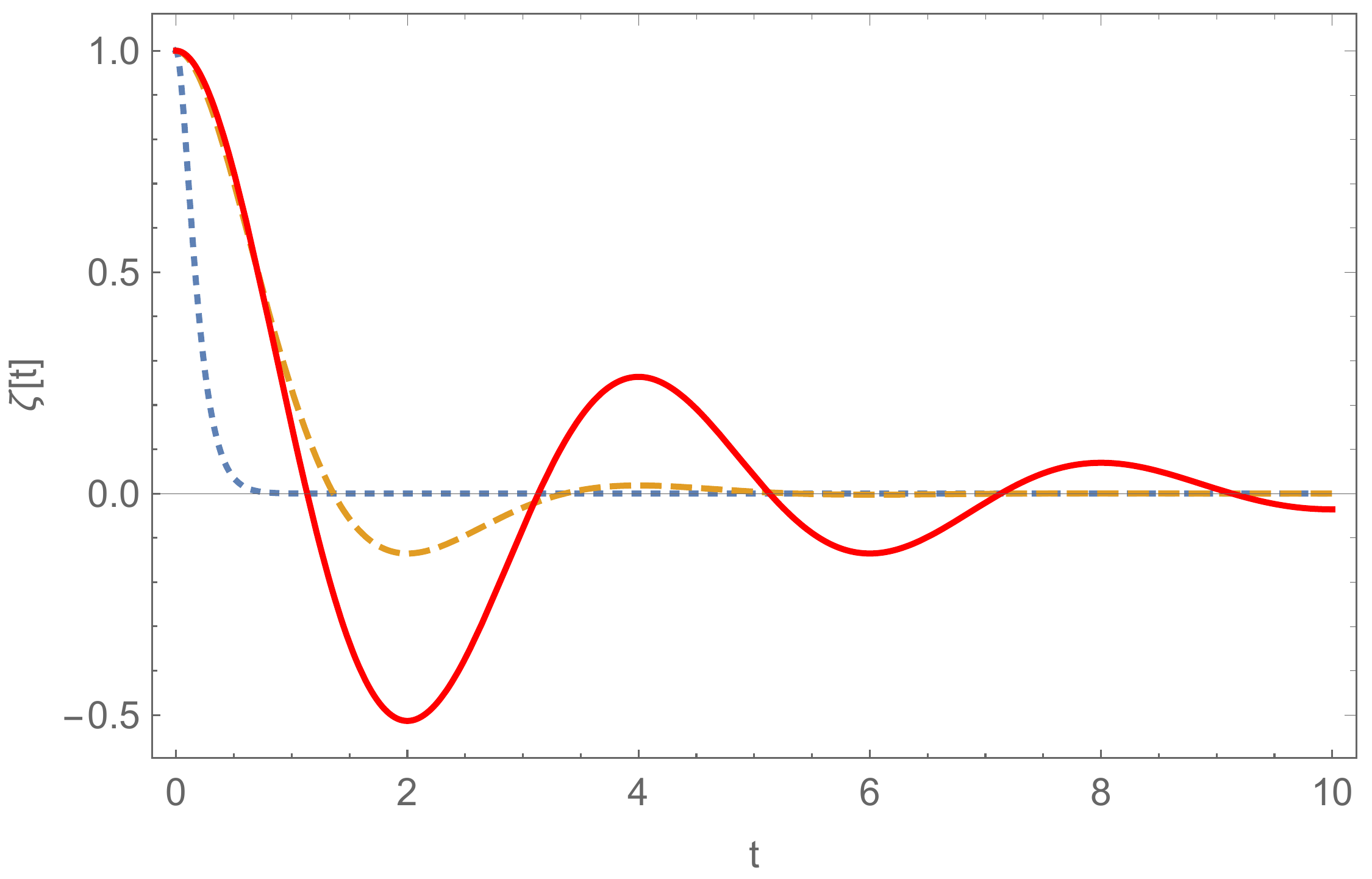}\\
  \caption{The friction kernel of the oscillating friction for different decay coefficient $\gamma$. In the plots, $\zeta=1$, and $\tilde{\omega}=\frac{\pi}{2}. $The dotted, dashed, and red lines are for $\gamma=0.1,1,3$.
  }\label{Oscillating_kernel}
\end{figure}

Our final model's friction kernel is given by
\begin{eqnarray}\label{osc_friction}
\zeta(t)=\zeta e^{-\frac{ |t|}{\gamma}}\left[\cos(\tilde{\omega}t)+\frac{1}{\gamma\tilde{\omega}}\sin(\tilde{\omega}|t|)\right]\;,
\end{eqnarray}
where $\zeta$ is the amplitude of the friction kernel denoting the strength of the friction, $\tilde{\omega}$ is the oscillating frequency of the friction kernel, and $\gamma$ is the decay coefficient or the time scale. It can be seen that the oscillatory friction kernel is the exponentially decayed friction plus the oscillating factor. In Figure \ref{Oscillating_kernel}, we have plotted the friction kernel of the oscillating friction for different decay coefficient $\gamma$. It can be observed that the friction kernel decays to zero in a finite time scale. So the analytical results are also valid for the oscillating friction kernel.

In order to calculate the MFPTs, we need the Laplace and Fourier transforms, which are given by  
\begin{eqnarray}
\hat{\zeta}(\lambda)=\frac{\zeta\left(\lambda+2/\gamma\right)}{\tilde{\omega}^2+\left(\lambda+1/\gamma\right)^2}\;,
\end{eqnarray}
and 
\begin{eqnarray}\label{Fourier_osc_friction}
\mathcal{F}(\omega)= \frac{4\gamma \zeta \left(\tilde{\omega}^2\gamma^2+1\right)}{\left(\left(\tilde{\omega}-\omega\right)^2\gamma^2+1\right)
\left(\left(\tilde{\omega}+\omega\right)^2\gamma^2+1\right)}\;.
\end{eqnarray}

From the analytical expressions of the MFPTs together with the Laplace and Fourier transforms of the oscillatory friction, it can be easily observed that there will also be kinetic turnover points when plotting the relation between the MFPT and the friction coefficient. Another observation is that the Fourier transforms of the oscillatory friction attain a maximum when $\omega=\tilde{\omega}$ due to the term $\left(\left(\tilde{\omega}-\omega\right)^2\gamma^2+1\right)$, which implies that in the weak friction regime, there will be a resonance in the kinetics when the oscillating frequency is equal to the oscillating frequency for the initial black hole in the potential well on the free energy landscape. The resonance effect will be more significant for large $\gamma$. In the following, we will pay more attention to the resonance effect in the kinetics of the black hole phase transitions.

\subsubsection{Dynamics of Hawking-Page phase transition}

\begin{figure}
  \centering
  \includegraphics[width=6cm]{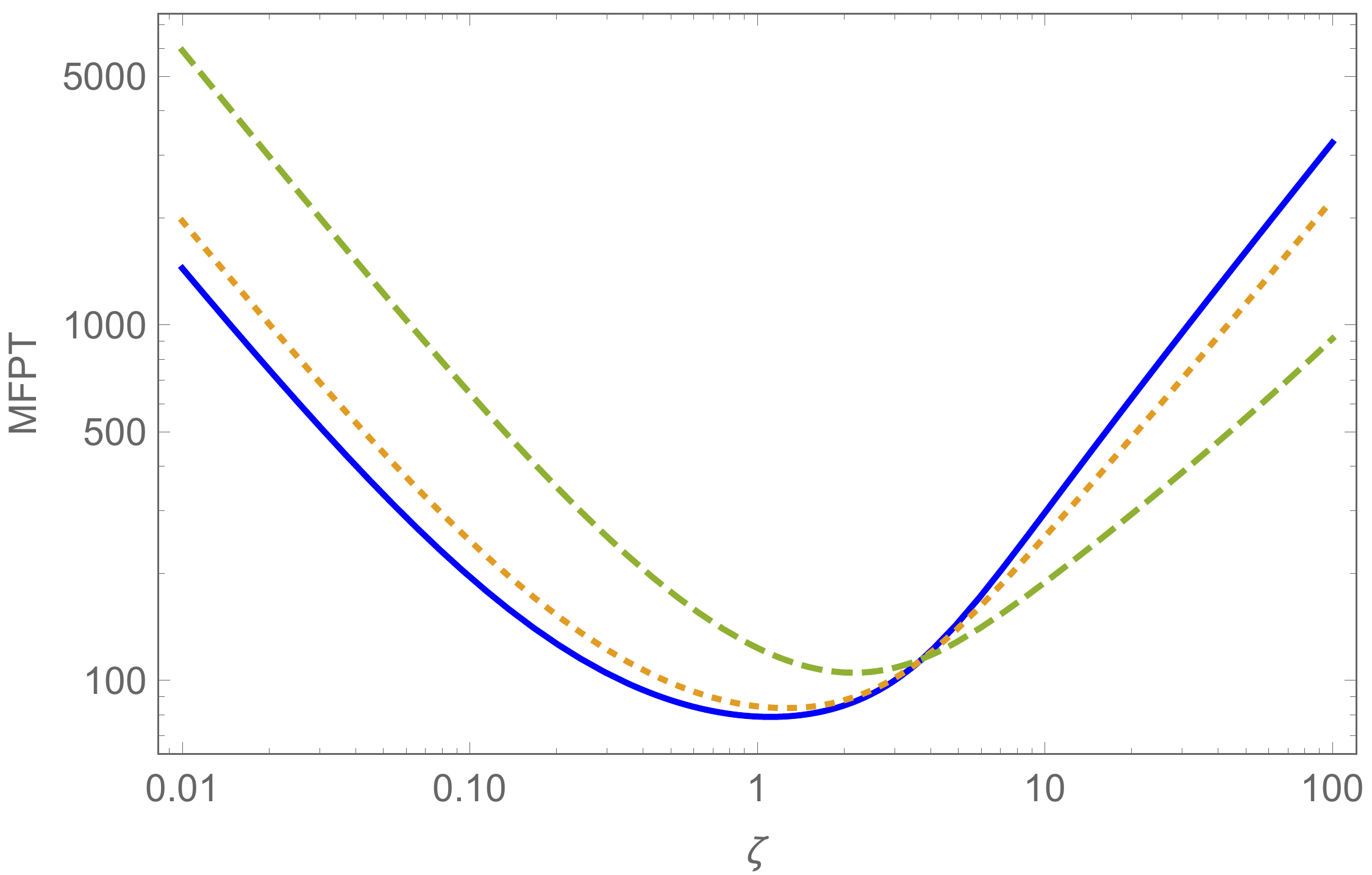}
  \includegraphics[width=6cm]{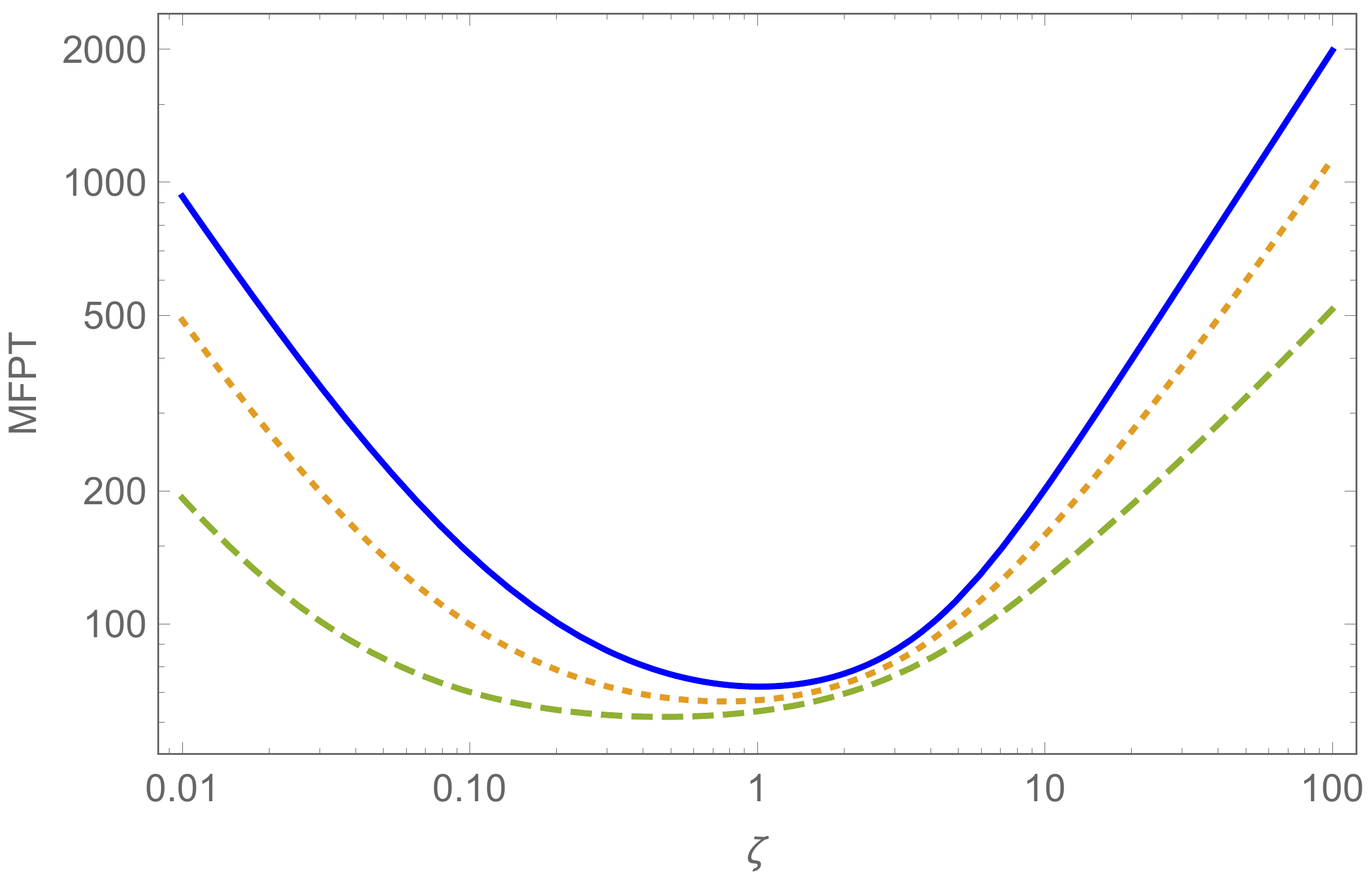}\\
  \caption{The dependency of the MFPT of the Hawking-Page phase transition from the large SAdS black hole state to the thermal AdS state on the friction coefficient for the oscillatory friction. In the plot, $L=1$, and $T=0.5$. The blue, dotted, and dashed lines are for $\gamma=1, 2.5, 10$. The left panel is for $\tilde{\omega}=\frac{\pi}{3}$ and the right panel is for $\tilde{\omega}=\frac{\pi}{2}$. 
  }\label{MFPTvsZeta_osc_HP}
\end{figure}

In Figure \ref{MFPTvsZeta_osc_HP}, we plot the dependency of the kinetic time of the Hawking-Page phase transition from the large SAdS black hole state to the thermal AdS state on the friction for the oscillatory friction kernel. There are kinetic turnover points for different $\gamma$ or $\tilde{\omega}$. The kinetic turnover is an universal property of black hole phase transition for both the Markovian model described by the time independent friction and the non-Markovian models described by the time dependent friction. The reason has been interpreted in terms of the analytical expressions.

We would like to discuss the non-Markovian effects on the black hole phase transition from the Figure \ref{MFPTvsZeta_osc_HP}. Note that for the oscillating friction given by Eq.(\ref{osc_friction}), the non-Markovian model can be reduced to the Markovian model in the $\gamma\rightarrow 0$ limit. Therefore, increasing the decay coefficient $\gamma$ means increasing the non-Markovian effects. For the left panel ($\tilde{\omega}=\frac{\pi}{3}$) in Figure \ref{MFPTvsZeta_osc_HP}, we can observe that in the weak friction regime, increasing $\gamma$ will slow down the transition process, and in the strong friction regime, increasing $\gamma$ will speed up the transition process. Therefore, for the oscillating frequency $\tilde{\omega}=\frac{\pi}{3}$, we have observed the same relation between the kinetics of black hole phase transition and the non-Markovian effects as in the last subsection. However, this conclusion is not valid in the right panel ($\tilde{\omega}=\frac{\pi}{2}$) in Figure \ref{MFPTvsZeta_osc_HP}. It can be seen that increasing the decay coefficient $\gamma$ will speed up the phase transition process in the whole friction regime. From Figure \ref{MFPTvsZeta_osc_HP}, We have observed that the non-Markovian effect may speed up or slow down the kinetics of the black hole phase transition, which depends on the oscillating frequency of the friction kernel. Therefore, we can conclude that due to the oscillating nature of the friction kernel, the non-Markovian effects on the kinetics of black hole phase transition can be complex and not universal for the oscillating friction kernel.

\begin{figure}
  \centering
  \includegraphics[width=6cm]{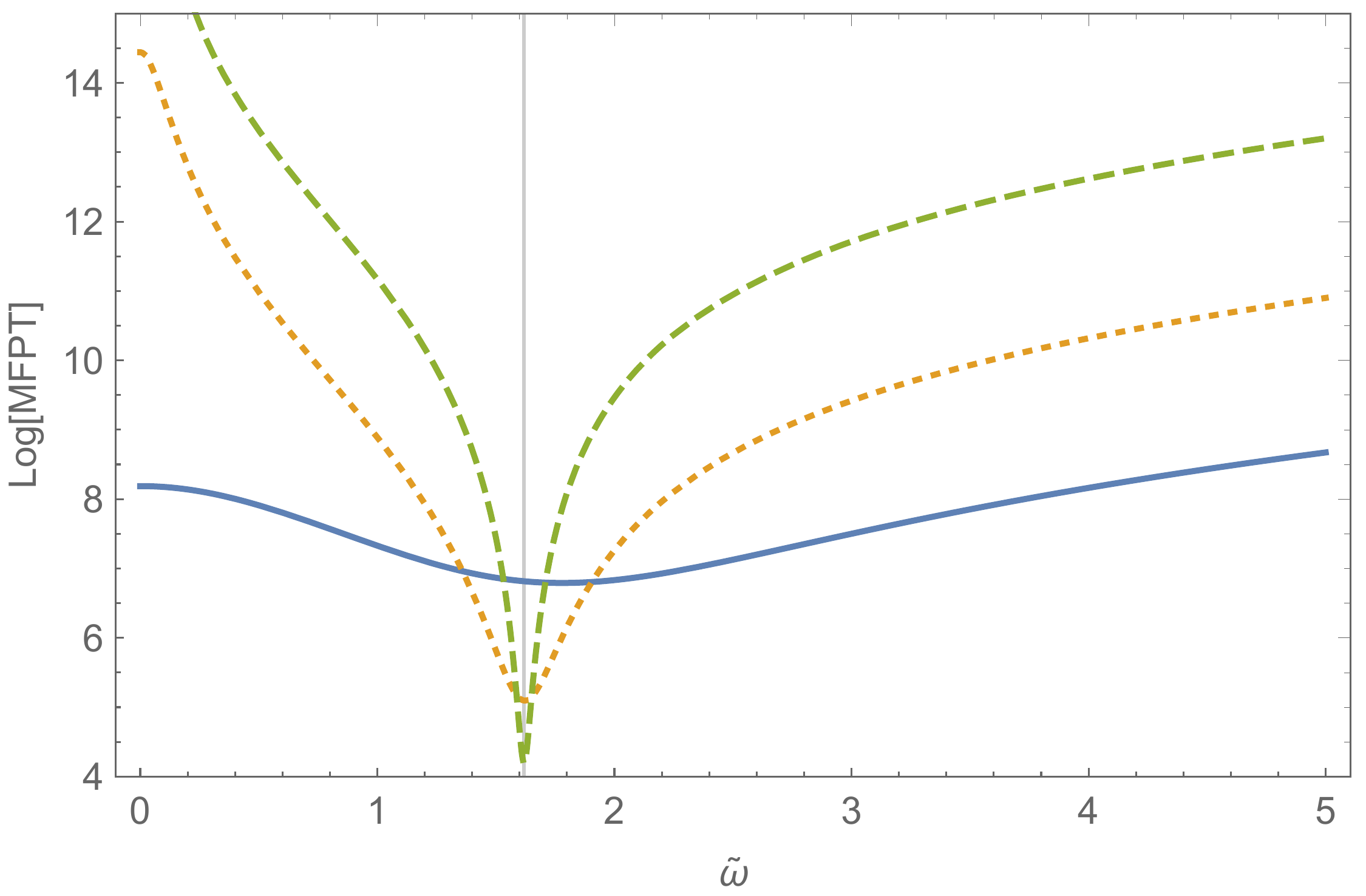}\\
  \caption{The MFPT of the Hawking-Page phase transition from the large SAdS black hole state to the thermal AdS state as a function of the oscillating frequency of the friction kernel. In the plot, $L=1$, $T=0.5$, and $\zeta=0.01$. The solid, dotted, and dashed lines are for $\gamma=1, 10, 100$. The vertical line represents $\tilde{\omega}=\omega_l$. 
  }\label{MFPTvsOmega_osc_HP}
\end{figure}

In Figure \ref{MFPTvsOmega_osc_HP}, we plot the MFPT of the Hawking-Page phase transition from the large SAdS black hole state to the thermal AdS state as a function of the oscillating frequency of the friction kernel. It can be observed that there is a resonance in the kinetics when the oscillating frequency is equal to the oscillating frequency for the initial black hole in the potential well on the free energy landscape. The reason has been explained below Eq.(\ref{Fourier_osc_friction}). Figure \ref{MFPTvsOmega_osc_HP} is plotted for the weak friction at $\zeta=0.01$. The resonance is not obvious at intermediate-strong friction regime. From the analytical expressions of the MFPTs in the weak friction regime together with the Fourier transforms of the oscillatory friction, we have $\bar{t}\sim \left(\left(\tilde{\omega}-\omega\right)^2\gamma^2+1\right)$. Therefore, there will be a minimum when $\tilde{\omega}=\omega$, In calculating the MFPT, $\omega$ is taken to be the oscillating frequency of the initial state in the initial potential well on the free energy landscape. For the Hawking-Page phase transition, $\omega=\omega_l$, where $\omega_l$ is the oscillating frequency of the right potential well on the free energy landscape of the Hawking-Page phase transition. When $\tilde{\omega}=\omega_l$, there is a resonance in the kinetics of transition process. It can also be observed that the kinetics resonates more clearly for the large decay coefficient or the time scale $\gamma$ that represents the non-Markovian effect.

\subsubsection{Dynamics of RNAdS black hole phase transition}

\begin{figure}
  \centering
  \includegraphics[width=6cm]{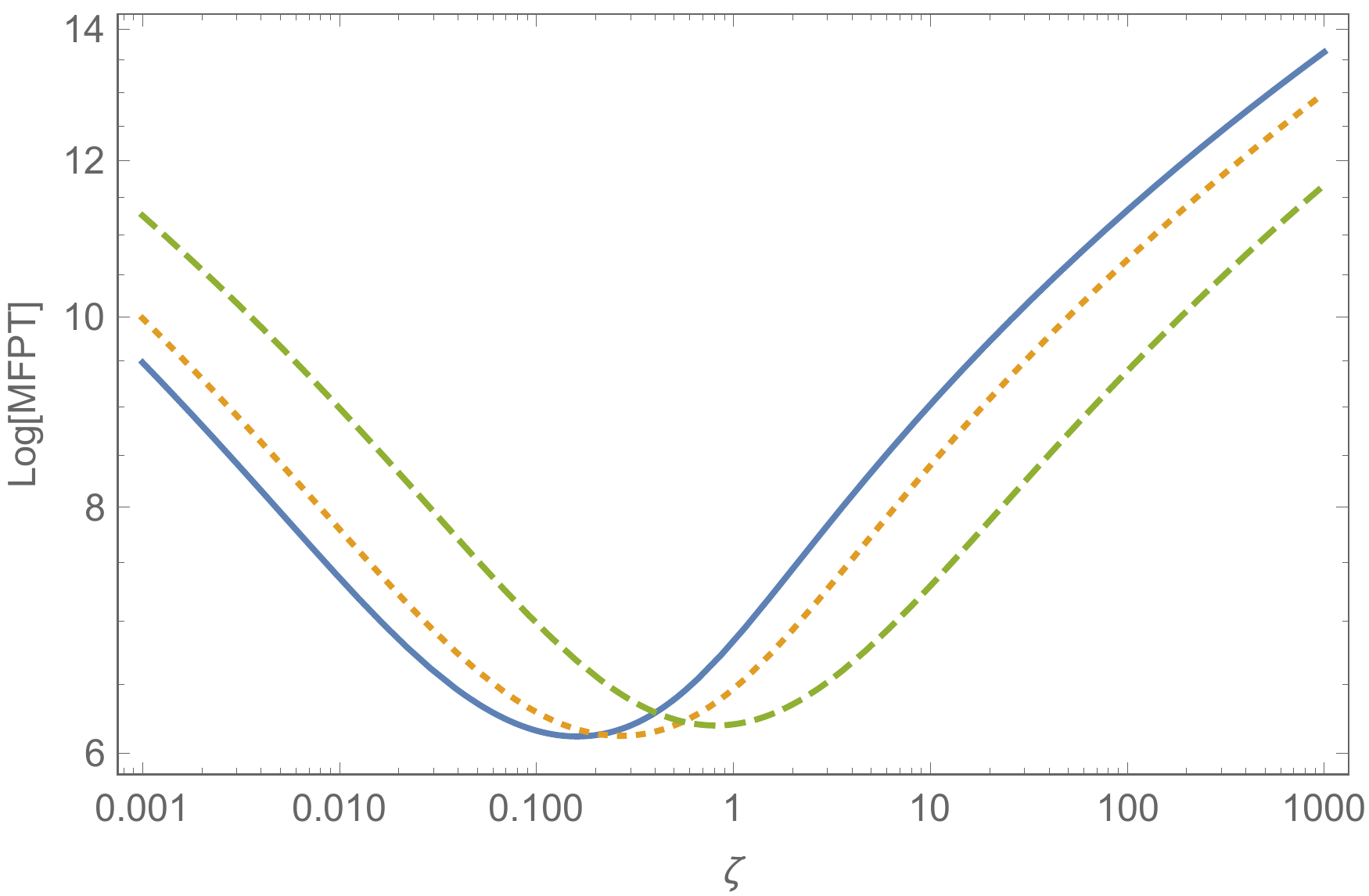}
  \includegraphics[width=6cm]{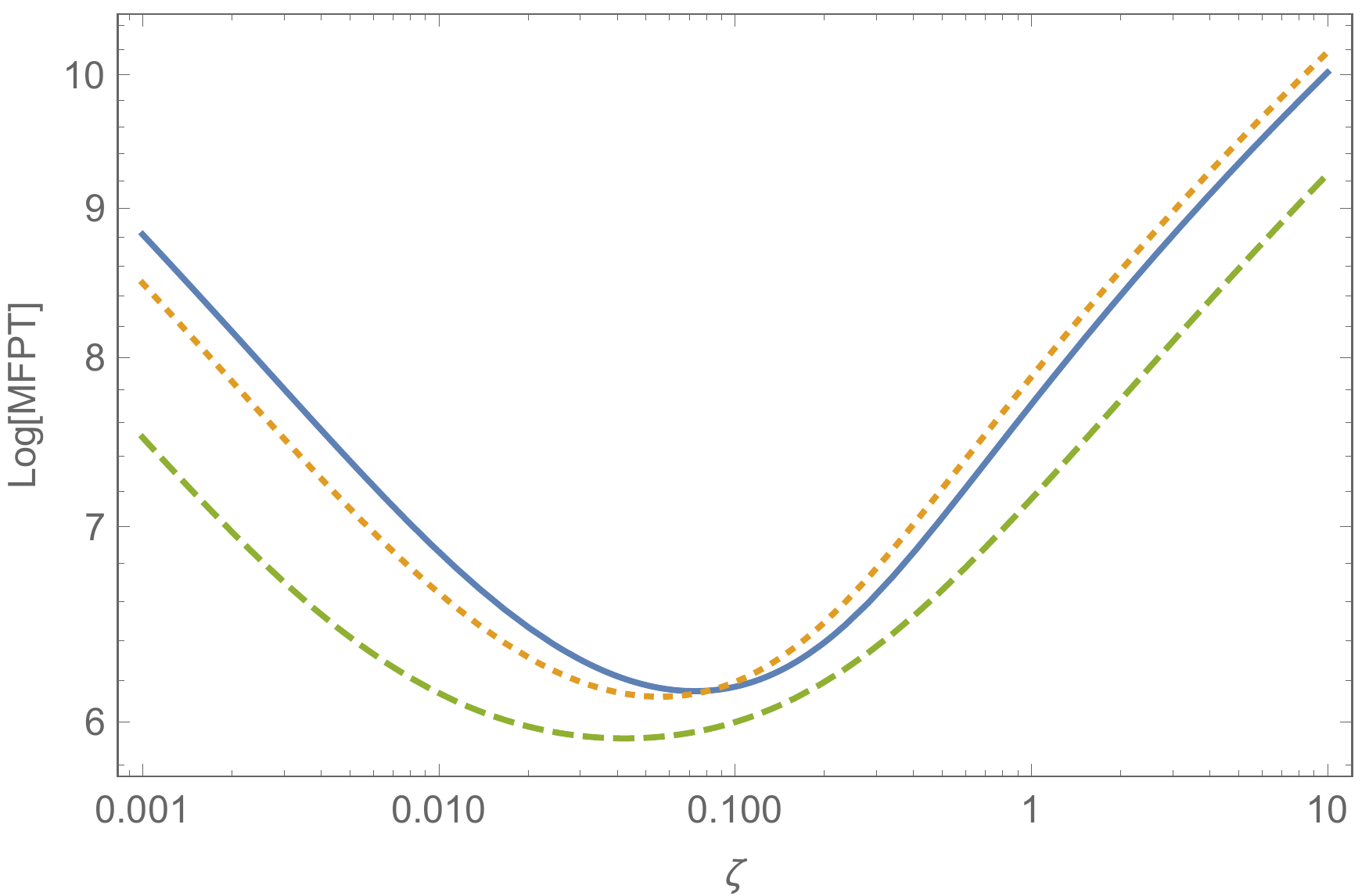}\\
  \caption{The MFPT of the RNAdS phase transition from the small black hole state to the large black hole state as a function of the friction coefficient for the oscillatory friction. In the plots, $Q=1$, $P=0.4P_c$, and $T=0.0298$. The , solid, dotted and dashed lines are for $\gamma=1, 2.5, 10$. The left panel is for $\tilde{\omega}=\frac{\pi}{2}$ and the right panel is for $\tilde{\omega}=\frac{\pi}{6}$.   }\label{MFPTvsZeta_osc_RNAdS_stl}
\end{figure}

In figure \ref{MFPTvsZeta_osc_RNAdS_stl}, we plot the MFPT of the RNAdS phase transition from the small black hole state to the large black hole state as a function of the friction coefficient for the oscillatory friction. Note that the MFPT as the function of the friction coefficient for the inverse process has the similar behavior. Once again, we observe that there are turnover points in the kinetic times when the friction coefficients change from the weak regime to the strong regime. As concluded from the numerical results for the Hawking-Page phase transition, for the oscillating friction kernel, there is no general rule for the dependency of the MFPT on the decay coefficient $\gamma$, i.e. there is no universal rule for the effects of the non-Markovian dynamics on the kinetics of the black hole phase transition.

\begin{figure}
  \centering
  \includegraphics[width=6cm]{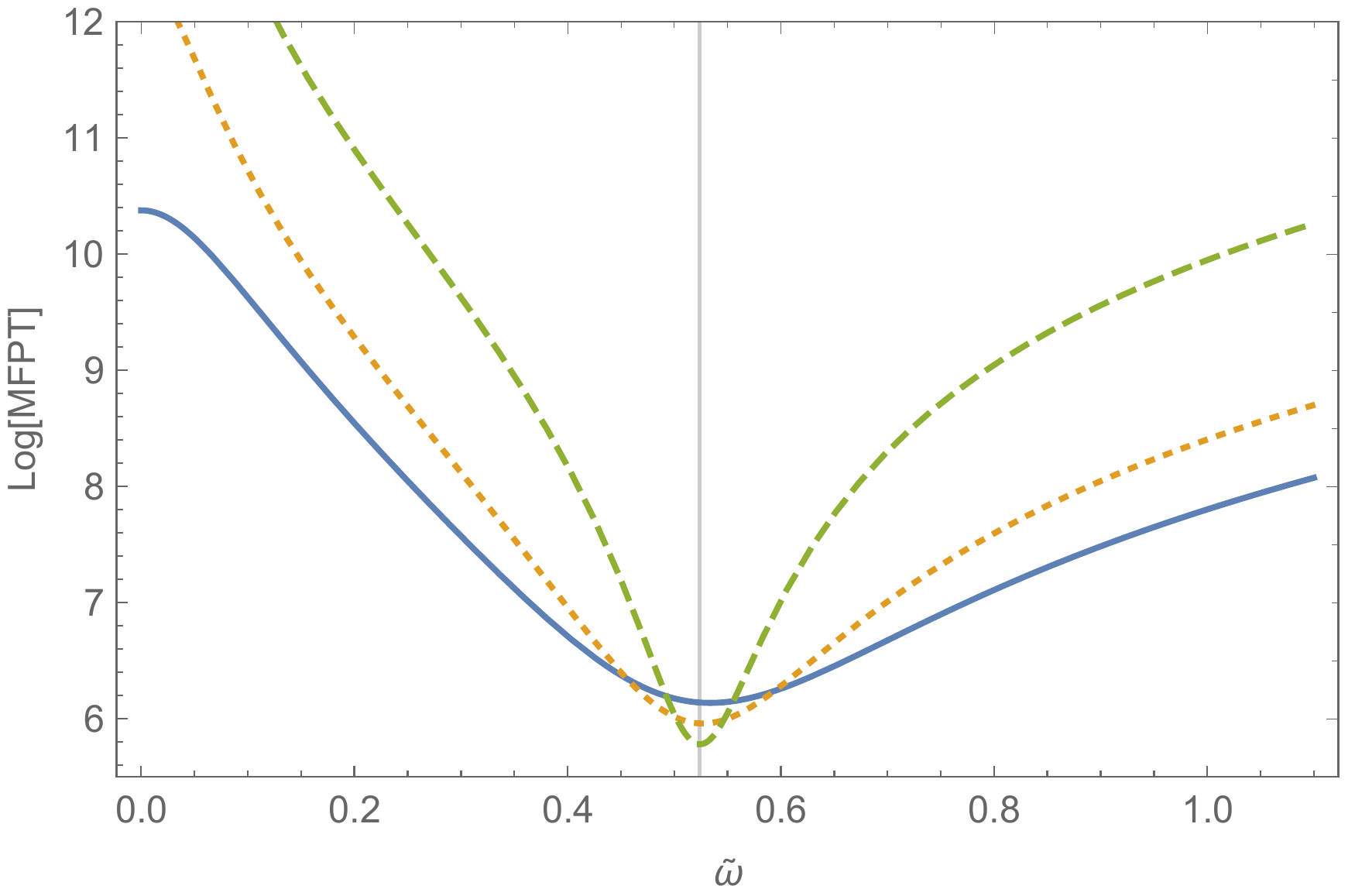}\\
  \caption{The MFPT of the RNAdS black hole phase transition from the small black hole state to the large one as a function of the oscillating frequency of the friction kernel. In the plot, $Q=1$, $P=0.4P_c$, $T=0.0298$, and $\zeta=0.01$. The solid, dotted, and dashed lines are for $\gamma=10, 20, 100$. The vertical line represents $\tilde{\omega}=\omega_s$. 
  }\label{MFPTvsOmega_osc_RNAdS}
\end{figure}

In Figure \ref{MFPTvsOmega_osc_RNAdS}, we plot the MFPT of the RNAdS black hole phase transition from the small black hole state to the large black hole state as a function of the oscillating frequency of the friction kernel. The results are plotted at fixed $\zeta=0.01$. There are kinetic resonances in the MFPTs for different decay coefficient $\gamma$ in the weak friction regime. In the intermediate-strong friction regime, the resonance in the kinetics is also not obvious for the RNAdS black hole phase transition. Because the initial state is selected to be the small RNAdS black hole state in the left potential well on the free energy landscape, the resonance appears when $\tilde{\omega}=\omega_s$, where $\omega_s$ is the oscillating frequency of the small black hole state.

\section{Conclusion and discussion}
\label{Con_dis}

In the present work, we performed the study of the non-Markovian effects on the black hole phase transition kinetics based on the free energy landscape. Free energy landscape allows us to not only analyze the thermodynamically favored state but also characterize the dynamical process of the transition from one state to another state. We have assumed that the effective stochastic dynamics describing the evolution along the order parameter of the black hole state is determined by the thermodynamic driving force, the effective friction along the order parameter, and the stochastic force from the effective heat bath form the underlying microscopic degrees of freedom of the black hole. The entire assumptions are at the macroscopic level emergent from the microscopic degrees of freedom. When the correlation time of the effective thermal bath is comparable or even longer than the oscillating time of the black hole state on the potential well of the free energy landscape, the non-Markovian model is required to study the kinetics of the black hole phase transition. In this case, the constant friction in the Markovian model should be replaced by the time dependent friction kernel. The stochastic evolution of the black hole order parameter is then governed by the generalized Langevin equation.

By treating the black hole phase transition process as stochastic, we considered the first passage problem and derived the analytical expressions of the MFPT in the weak, intermediate, and large friction regimes. From these analytical expressions, we can anticipate that the kinetic turnover of the black hole phase transition is an universal property not only for the Markovian dynamics but also for the non-Markovian dynamics. As the concrete examples, we study the effects of three kinds of time dependent friction kernels (i.e., delta function like, exponentially decayed, and oscillatory decayed frictions) on the kinetics of Hawking-Page phase transition and the phase transitions of the small/large RNAdS black holes. The numerical results show that there are kinetic turnover when the friction strength changes from the weak regime to the strong regime. Another interesting observation is that barrier height on the free energy landscape is the dominant contribution to the kinetic time.

For the delta friction kernel, the model is reduced to the Markovian model. In fact, the kinetics of the RNAdS black hole phase transition has been studied numerically in the previous work \cite{Li:2021vdp}. As a complement, the kinetics of the Hawking-Page phase transition in the whole friction regime is studied by using the analytical method in this work. For the exponentially decayed friction kernel, we mainly focus on the non-Markovian effects. It is shown that the non-Markovian effect slows down the phase transition dynamics in the weak friction regime and speeds up the transition process in the strong friction regime. For the oscillating decayed friction kernel, we consider the effects of the oscillating frequency of the effective thermal bath on the kinetics. It is shown that there are kinetic resonances when the oscillating frequency of the effective thermal bath is equal to the oscillating frequency of the black hole state in the initial potential well on the free energy landscape.

At last, we discuss two possible aspects that deserve further investigation. In this work, we only discuss two types of black hole phase transitions. One can generalize the current work to study other phase transition processes in black hole physics. The analytical method is employed to study the non-Markovian dynamics of the black hole phase transition. Note that the generalized Langevin equation is an integro-differential equation. Solving the generalized Langevin equation numerically will give the exact results of the kinetics of the black hole phase transition, which is still challenging at the moment and requires further investigation.

\end{document}